%% file: main.tex
\documentclass[11pt]{article}
\usepackage{amssymb}
\usepackage{graphicx}
\usepackage{chngcntr}
\usepackage{import}

\usepackage{longtable}
\usepackage{lscape}

\newcommand{\deltanu}{\mbox{$\langle \Delta\nu \rangle$}}
\newcommand{\numax}{\mbox{$\nu_{\rm max}$}}

\def\aj{{\textit{Astron. J.}}}                   
\def\araa{{\textit{Annu. Rev. Astron. Astrophys.}}}             
\def\apj{{\textit{Astrophys. J.}}}                 
\def\apjl{{\textit{Astrophys. J.}}}                
\def\apjs{{\textit{Astrophys. J. Suppl. Ser.}}}               
\def\apss{{\textit{Astrophys.~Space~Sci.}}}             
\def\aap{{\textit{Astron. Astrophys.}}}                
          
\def\mnras{{\textit{Mon.~Not.~R.~Astron.~Soc.}}}

\def\pasp{{\textit{Publ.~Astron.~Soc.~Pac.}}}

\def\nat{{\textit{Nature}}}

\begin{document}
	

\vspace*{5mm}

\begin{center}
\textbf{\Large{Chronologically dating the early assembly of the Milky Way}}\\
\end{center}

\vspace{5mm}


\noindent Josefina Montalb\'an$^{1}$, J. Ted Mackereth$^{1,2,3,4}$, Andrea Miglio$^{1,5,6}$, Fiorenzo Vincenzo$^{1,7,8}$, Cristina Chiappini$^{9}$, Gael Buldgen$^{10}$, Beno\^it Mosser$^{11}$, Arlette Noels$^{12}$, Richard Scuflaire$^{12}$, Mathieu Vrard$^{8,13}$, Emma Willett$^{1,14}$, Guy R. Davies$^{1,14}$, Oliver Hall$^{1,14}$, Martin Bo Nielsen$^{1,14,15}$, Saniya Khan$^{1,14}$, Ben M. Rendle$^{1,14}$, Walter E. van Rossem$^{1,14}$, Jason W. Ferguson$^{16}$, William J. Chaplin$^{1,14}$\\

\vspace{5mm}

\small{
\noindent  $^{1}$ School of Physics and Astronomy, University of Birmingham, Birmingham B15 2TT, UK

\noindent  $^{2}$ Department of Astronomy and Astrophysics, University of Toronto, 50 St. George Street, Toronto ON, M5S 3H4, Canada

\noindent  $^{3}$ Canadian Institute for Theoretical Astrophysics, University of Toronto, 60 St. George Street, Toronto ON, M5S 3H8, Canada

\noindent  $^{4}$ Dunlap Institute for Astronomy and Astrophysics, University of Toronto, 50 St. George Street, Toronto ON, M5S 3H4, Canada

\noindent  $^{5}$ Dipartimento di Fisica e Astronomia, Universit{\`a} degli Studi di Bologna, Via Gobetti 93/2, I-40129 Bologna, Italy

\noindent  $^{6}$ INAF -- Astrophysics and Space Science Observatory Bologna, Via Gobetti 93/3, I-40129 Bologna, Italy

\noindent  $^{7}$ Center for Cosmology and AstroParticle Physics, The Ohio State University, 191 West Woodruff Avenue, Columbus, OH 43210, USA

\noindent  $^{8}$ Department of Astronomy, The Ohio State University, Columbus, OH 43210, USA

\noindent  $^{9}$ Leibniz-Institut f\"ur Astrophysik Potsdam (AIP),  An der Sternwarte 16, 14482 Potsdam, Germany

\noindent  $^{10}$ Observatoire de Gen\`eve, Universit\'e de Gen\`eve, 51 Ch. Des Maillettes, 1290 Sauverny, Switzerland

\noindent  $^{11}$ LESIA, Observatoire de Paris, Universit\'e PSL, CNRS, Sorbonne Universit\'e, Universit\'e de Paris, 5 place Jules Janssen, 92195 Meudon, France

\noindent  $^{12}$ Space Sciences, Technologies and Astrophysics Research (STAR) Institute, Universit\'e de Li\`ege, 19c All\'ee du 6 Ao\^ut, B-4000 Li\`ege, Belgium

\noindent  $^{13}$ Instituto de Astrof\'isica e Ci\^encias do Espa\c{c}o, Universidade do Porto, CAUP, Rua das Estrelas, 4150-762 Porto, Portugal

\noindent  $^{14}$ Stellar Astrophysics Centre (SAC), Department of Physics and Astronomy, Aarhus University, Ny Munkegade 120, DK-8000 Aarhus C, Denmark

\noindent  $^{15}$ Center for Space Science, NYUAD, New York University Abu Dhabi, PO Box 129188, Abu Dhabi, United Arab Emirates

\noindent  $^{16}$ Department of Physics, Wichita State University, Wichita, KS 67260-0032, USA

}

\vspace{15mm}

\textbf{The standard cosmological model ($\Lambda$-CDM) predicts that galaxies are built through hierarchical assembly on cosmological timescales\cite{Peebles1993,Kauffmann1993}. The Milky Way, like other disc galaxies,  underwent violent mergers and accretion of small satellite galaxies in its early history. Thanks to Gaia-DR2\cite{gaia} and spectroscopic surveys\cite{apogee}, the stellar remnants of such mergers have been identified\cite{Helmi2018,Belokurov2018,Myeong2019}. The chronological dating of such events is crucial to uncover the formation and evolution of the Galaxy at high redshift, but it has so far been challenging owing to difficulties in obtaining precise ages for these oldest stars. Here we combine asteroseismology -- the study of stellar oscillations -- with kinematics and chemical abundances, to estimate precise stellar ages ($\sim$~11\%) for a sample of stars observed by the \emph{Kepler} space mission\cite{Borucki2016}. Crucially, this sample includes not only some of the oldest stars that were formed inside the Galaxy, but also stars formed externally and subsequently accreted onto the Milky Way. Leveraging this resolution in age, we provide compelling evidence in favour of models in which the Galaxy had already formed a substantial population of its stars (which now reside mainly in its thick disc) before the in-fall of the satellite galaxy Gaia-Enceladus/Sausage\cite{Helmi2018,Belokurov2018} around 10 billions years ago.}\\


Recent results based on the ESA Gaia mission\cite{gaia} have revealed that the stellar content of the inner halo of the Milky Way (MW) is dominated by debris from some seemingly massive dwarf galaxies, such as the Gaia-Enceladus/Sausage (hereafter GES).The merging event with the GES is now purported to be one of the most important in the Galaxy's history, shaping how we observe it today\cite{Belokurov2018, Belokurov2019, DiMatteo2019, Vincenzo2019}. To constrain the effect of such mergers on the MW and other similar galaxies, it is crucial to understand their state both prior to and following the merger. This requires mapping the temporal sequence of these events with the highest precision possible ($\sim 10$\% to follow the first 4 billion years after the Big Bang\cite{Miglio2017}). Several recent works have estimated the characteristics and timing of this merging event \cite{Helmi2018,Haywood2018,DiMatteo2019,Mackereth2019,Vincenzo2019,Grand2020}, while others (before\cite{Schuster2012,Hawkins2014} and after\cite{Das2020,Gallart2019} {Gaia}-DR2) have aimed to age date the accreted and \emph{in-situ} stellar populations of the MW halo (see Helmi\cite{Helmi2020} for an extensive review). Although using different kind of targets and methods, these age dating techniques are however quite limited in precision and accuracy, since they are based on stellar surface properties and on predictions from stellar evolution models. The latter are known to be affected by e.g. uncertainty in the physics and degeneracy between parameters, which makes it difficult to obtain stellar ages with the required precision and accuracy.

Red giant (RG) stars, being long-lived and intrinsically bright, are excellent candidates to map ages in different regions of the MW\cite{Miglio2013,Miglio2017,Das2020}. However, ageing RGs in color-magnitude space using their surface properties gives yet uncertain results since their colours and luminosities are similar, whatever their mass and age. Fortunately,  asteroseismology, which probes the internal structure of stars, provides us with means to reach a  precision  of 10-20\% on age-dating individual RG stars\cite{ChaplinMiglio2013,Miglio2017,Miglio2020}.
 
Among the roughly 15,000 oscillating K- and G-type RG stars detected in the field observed by the NASA \emph{Kepler} space telescope\cite{Borucki2016}, a small fraction lie in the low-metallicity regime characteristic of the inner MW halo and high-$\mathrm{[\alpha/Fe]}$ disc component ($\mathrm{[Fe/H]} <-0.5$). Of these, some 400 stars have precisely measured element abundances, atmospheric  parameters and radial velocities from the Data Release 14 of the Apache Point Observatory Galactic Evolution spectroscopic survey  (APOGEE DR14\cite{apogee}), as well as detailed proper motions from Gaia-DR2\cite{gaia}. 


\begin{figure}
	\centering
	\includegraphics[width=0.9\textwidth]{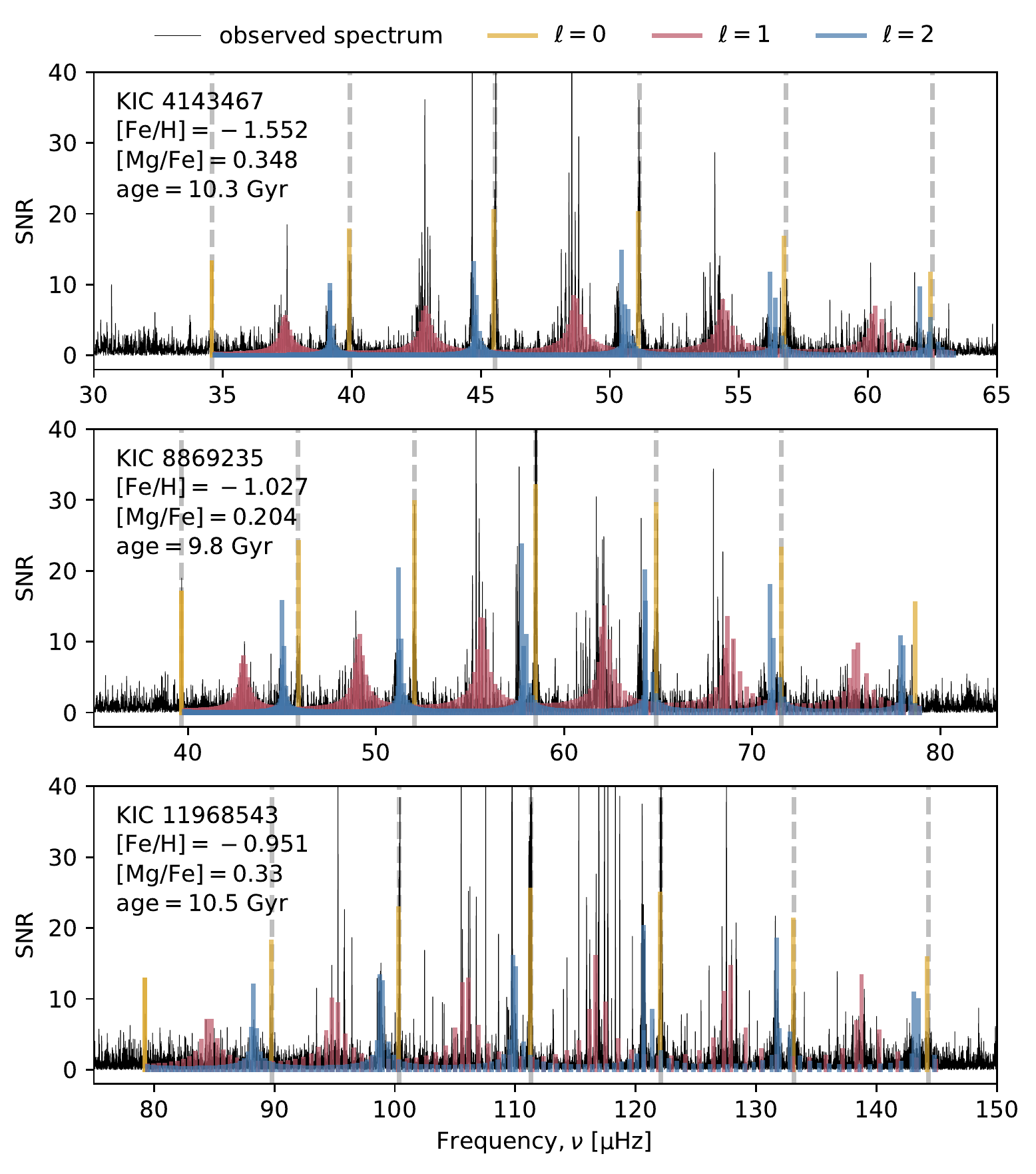}
	
	\caption{\small {\bf Observed and modelled stellar power spectra.} The power spectra of three stars from our sample which span a wide range in $\mathrm{[Fe/H]}$, compared with the theoretical spectra of best fit model returned by AIMS. SNR (signal-to-noise ratio) represents the height of the mode peaks relative to the surrounding noise floor. The vertical lines, coloured by their angular degree $\ell$, show the frequency and relative estimate amplitude of the oscillation modes. Only the measured $l=0$ mode frequencies (shown by gray dashed vertical lines) were used as seismic constraints in AIMS. The $\ell=1$ and $\ell=2$ mode frequencies predicted by the best fit model reproduce those modes visible in the data with good accuracy, reinforcing the confidence on the quality of the fitting procedure and of the derived stellar parameters.}
	
	\label{fig:powspec}
\end{figure}


The \emph{Kepler} data provide oscillation frequency spectra of exquisite quality and resolution, allowing precise estimates to be made of the frequencies of modes of different angular degree (radial $\ell=0$,  dipolar $\ell=1$ and quadrupolar $\ell=2$), and hence of fundamental parameters and evolutionary state of the stars. We first use this seismic information to remove from the sample those stars that are in the red clump (RC) phase (i.e. low-mass He-core-burning stars that have likely underwent mass loss earlier in their evolution) and in subsequent phases. Removing these contaminants leaves a sample of $95$ red giant branch (RGB) stars whose radial-mode frequencies we have measured to use as the asteroseismic input for inferring robust and precise ages (see  \textbf{Methods} and Extended Data Fig.~5 for details).


\begin{figure*}
	\includegraphics[width=\linewidth]{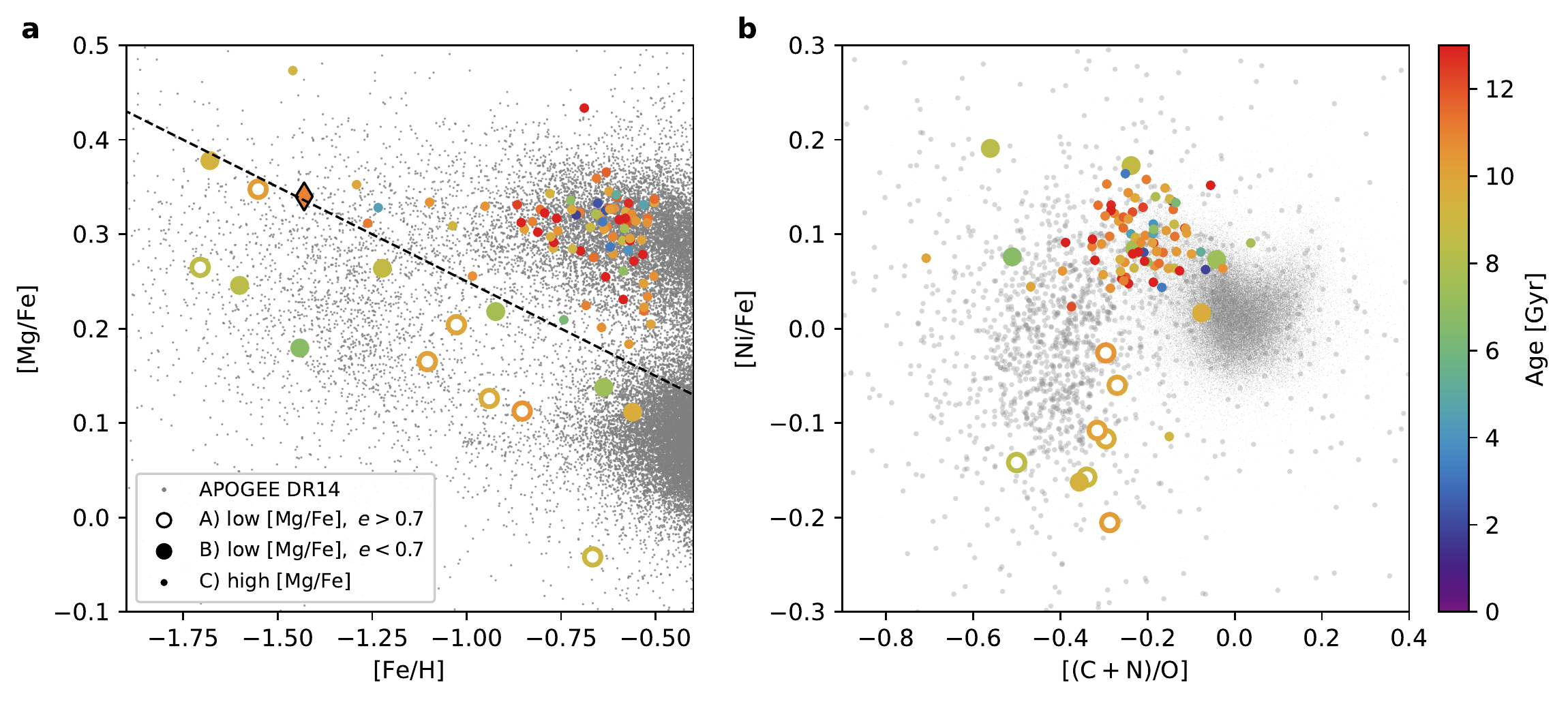}
	
	\caption{\small {\bf Chronological, chemical and kinematic properties of the seismic RGB sample.} {\bf a}, The $\mathrm{[Mg/Fe]}$-$\mathrm{[Fe/H]}$ plane for our sample (points coloured by age), compared with the rest of APOGEE DR14 at $\mathrm{[Fe/H]}< -0.5$ (small grey points). The dashed line ($\mathrm{[Mg/Fe]} = -0.2\ \mathrm{[Fe/H]} + 0.05$) demonstrates the simple division we make between likely \emph{in-situ} (above) and \emph{ex-situ} (below) stars. Likely \emph{ex-situ} stars with $e > 0.7$ are shown as open points. The diamond represents $\nu$~Indi values\cite{Chaplin2020}. {\bf b}, The $\mathrm{[(C+N)/O]}$-$\mathrm{[Ni/Fe]}$ distribution of the three groups  defined in panel {\bf a}. The underlying gray points represent the entire APOGEE DR14 sample with $\mathrm{[Fe/H]} < -0.5$, with those which lie below the $\mathrm{[Fe/H]}$-$\mathrm{[Mg/Fe]}$ division shown as larger points. The low-$\mathrm{[Mg/Fe]}$,  $e > 0.7$ stars have  atypical element abundances relative to the other groups, exhibiting very low $\mathrm{[Ni/Fe]}$ and a small depletion in Carbon and Nitrogen relative to Oxygen.}
	
	\label{fig:selection}
\end{figure*}

We estimate stellar properties (Supplementary Table.~1) using the individual frequencies of radial modes and atmospheric parameters from spectroscopy as observational inputs in AIMS\cite{aims}, a Bayesian parameter estimation code, which provides best-fitting stellar properties and full posterior probability distributions by comparing with theoretical stellar models and adiabatic frequencies (see \textbf{Methods} and Extended Data Fig.~6,~7). The precision on age we achieve, of 11\% on average, affords us the ability to unpick the chronological sequence of events some $\sim 12$ Gyr ago, as we show below.

The robustness of our estimated stellar age distributions has been checked performing  different tests (see \textbf{Methods} section and Extended Data Figs.~8,~9, for detailed description). Moreover, as shown in Fig.~1, despite having only fitted modes of degree $\ell=0$, the theoretical spectra predicted by the best-fitting model parameters reproduce well also the non-radial modes ($\ell=1,$ 2) of the observed spectra, which reinforces the confidence on the quality of the derived stellar parameters. 


\begin{figure}
	\centering
	\includegraphics[width=0.725\linewidth]{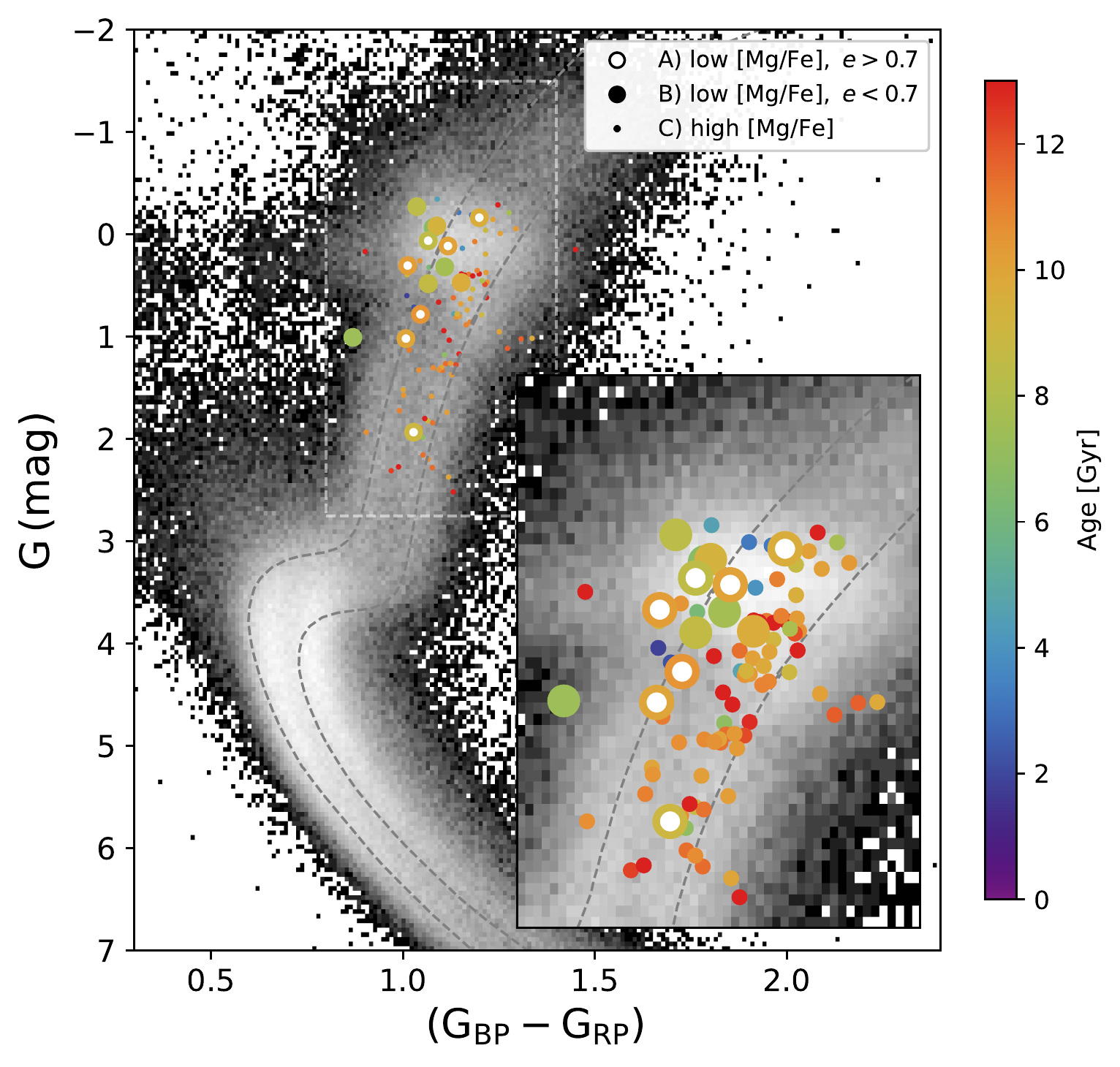}
	
	\caption{\small\textbf {Gaia-DR2 colour-magnitude diagram (CMD) for our sample and kinematically defined halo.} 2D gray histogram represents the stars with tangential velocity $\mathrm{v_{T}} > 200 \,\mathrm{km\,s^{-1}}$, and age-scaled coloured symbols (same as in Fig.~2) represent our 95 RGB smaple.   This CMD (absolute magnitude in the Gaia passband  – G – versus color represented by the difference of the Gaia magnitudes in the Gaia BP and RP passbands) has been already discussed\cite{Gallart2019,Babusiaux2016,Haywood2018}, but we show here that the stars we identify as members of the \emph{ex-situ} halo (group A) lie mainly on the blue sequence and have, on average, younger seismic ages than the stars of group C (\emph{in-situ} stars) on the red sequence. For the sole purpose of guiding the eye, we have added two isochrones\cite{PARSEC} (grey dashed lines) following to the blue and red sequences.}
	
	\label{fig:cmd}
\end{figure}


Figure~2 summarises the chronological, chemical and kinematic properties of the final sample of 95 RGB stars for which we could robustly determine ages. In Fig.~2a we show its $\mathrm{[Fe/H]}$-$\mathrm{[Mg/Fe]}$ distribution (coloured by age), together with that of APOGEE DR14 sample (grey points). The grey points clearly show, in addition to two over-densities at higher $\mathrm{[Fe/H]}$ corresponding to the low- and high-$\mathrm{[\alpha/Fe]}$ Galactic disc populations and  metal-rich \emph{in-situ} halo\cite{Hayden2015}, a scattered population at $\mathrm{[Fe/H] < -0.7}$ and intermediate $\mathrm{[Mg/Fe]}$ ($\sim 0.1$ to $0.2$), where the recently characterised GES population lies\cite{Helmi2018,Hayes2018,Mackereth2019}.  Our final RGB sample (coloured circles) contains members of each of these populations. We also include, for reference, the location of $\nu$~Indi, a bright sub-giant recently dated using seismology, and classified as belonging to the ``heated" thick disc\cite{Chaplin2020}.

Recent studies of the Galactic halo and local group dwarfs\cite{Tolstoy2009} suggest that stars in the low-$\mathrm{[Mg/Fe]}$ sequence at $\mathrm{[Fe/H] \lesssim -0.7}$  have likely been accreted to the Galaxy\cite{NissenSchuster2010, Schuster2012, Hawkins2015, Fernandez-Alvar2018, Hayes2018}.  The $\mathrm{[\alpha/Fe]}$ ratios in local dwarfs indicate a higher pollution from Type Ia Supernovae relative to core-collapse SNe,  likely due to inefficient star-formation activity and strong outflows\cite{Salvadori2009,Vincenzo2014}. As a consequence,  their $\mathrm{[\alpha/Fe]}$ ratios are lower than \emph{in-situ} halo stars, where element abundances are more affected by nucleosynthetic products from core-collapse as opposed to Type Ia Supernovae. 

Based on the above studies, we classify the asteroseismic RG sample by making a cut in $\mathrm{[Fe/H]}$-$\mathrm{[Mg/Fe]}$ space along the line $\mathrm{[Mg/Fe]} = -0.2\ \mathrm{[Fe/H]} + 0.05$. Stars below this line are likely to have been formed in dwarf satellites and then accreted, and those above should be born, in majority, \emph{in-situ}. It is conceivable that the \emph{in-} and \emph{ex-situ} populations defined in this way will have some contamination from the other group. To mitigate this, we further divide stars below the line into high and low orbital eccentricity groups (calculated as described in \textbf{Methods}). Stars on more radial orbits (eccentricities $e > 0.7$, open points) are those most likely to have been accreted from the GES progenitor\cite{Mackereth2019}.

Figure~2b, which shows the Nickel abundance relative to Iron $\mathrm{[Ni/Fe]}$ and the sum of Carbon and Nitrogen abundance relative to Oxygen $\mathrm{[(C+N)/O]}$, supports that the applied cuts efficiently isolate different stellar populations\cite{NissenSchuster2010, Hawkins2015}. The APOGEE-DR14 sample below the $\mathrm{[Fe/H]}$-$\mathrm{[Mg/Fe]}$ line (large grey points) is depleted in both $\mathrm{[Ni/Fe]}$ and $\mathrm{[(C+N)/O]}$, consistent with local dwarf satellite galaxies which contain stars with $\mathrm{[Ni/Fe]}$ ratios  lower than the MW\cite{McWilliam2018,Kirby2019}. The stars of our high-$e$, low-$\mathrm{[Mg/Fe]}$ sample [hereafter group (A)] lie at the lowest values of $\mathrm{[Ni/Fe]}$, and are clearly separated from the other groups, reinforcing  our contention that the low-$\mathrm{[Mg/Fe]}$, $e > 0.7$ group is likely to be formed \emph{ex-situ}. The group made of low-$e$, low-$\mathrm{[Mg/Fe]}$ stars [hereafter group (B)], is probably a mixture of stars of different origin \cite{Mackereth2019, Koppelman2019}: the tail at low eccentricity of GES stars, the low-metallicity end of the thin disc (for instance,  the two stars with  $\mathrm{[Fe/H]}\gtrsim-0.7$), or remnants  of less massive accretion events,  as could be  the case for the stars with $\mathrm{[Ni/Fe]}$ and $\mathrm{[(C+N)/O]}$ patterns similar to those of  high-$\mathrm{[Mg/Fe]}$ stars, which could be indicative of a different star formation history in the galaxy of origin\cite{McWilliam2018,Kirby2019}. The composition of this group is the most sensitive to the details of the classification criterion adopted, however, this does not affect the robustness of our main conclusions  about the chronological order of the GES and high-$\mathrm{[Mg/Fe]}$ [hereafter group (C)] populations (see section {\textbf {Methods}}, Extended Data Fig.~10).

\begin{figure}
	\centering
	\includegraphics[width=0.75\textwidth]{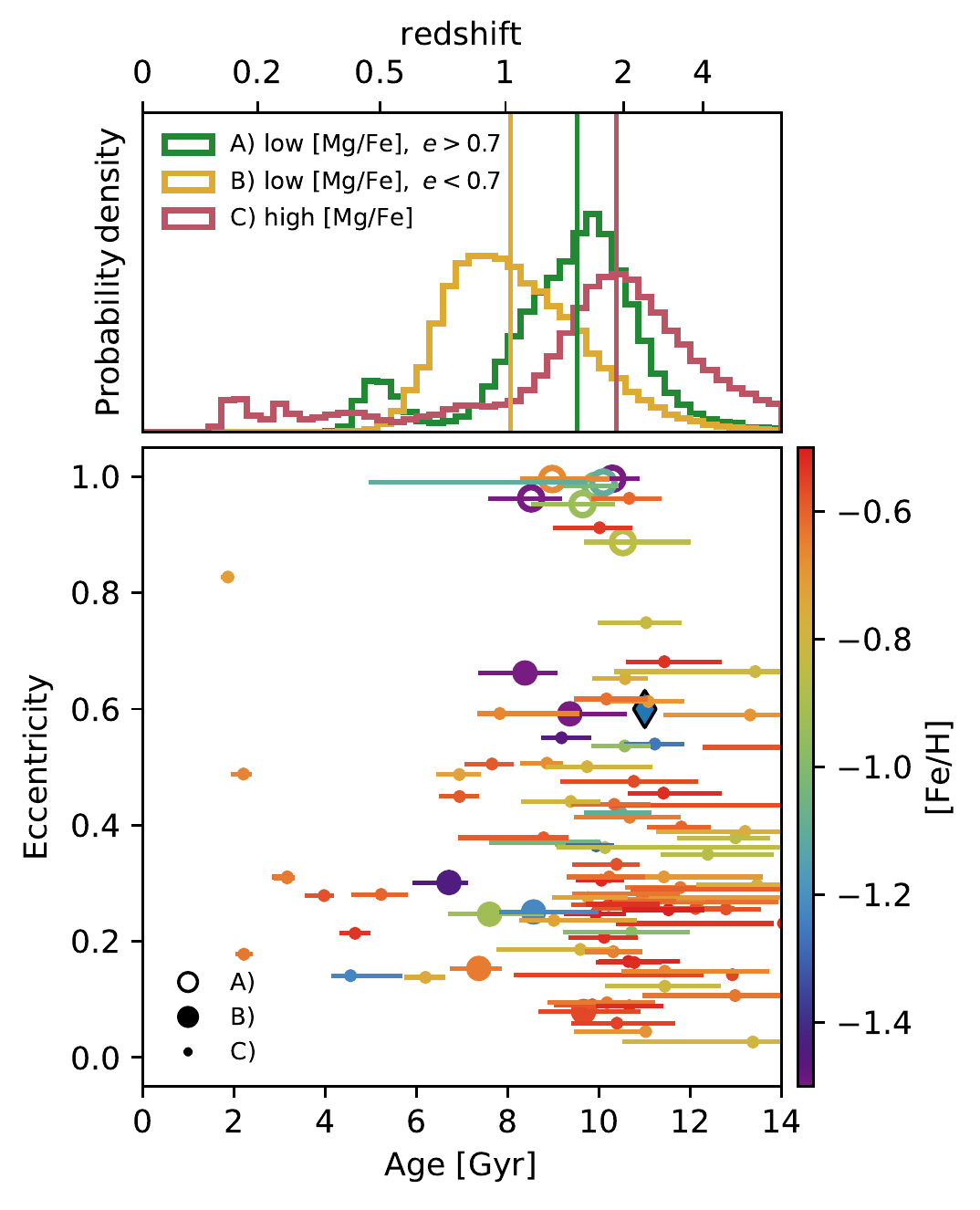}
	
	\caption{\small {\bf Age and eccentricity distributions.} Age against eccentricity ($e$) for the stars in the sample coloured by $\mathrm{[Fe/H]}$. Circles represent the age values of the best fitting models, and horizontal lines their uncertainties ([16\%-84\%] credible interval from full posterior distributions). Uncertainties on $e$ are smaller than the symbol size. The diamond represents $\nu$~Indi\cite{Chaplin2020} (not included in the distributions). The histogram above reflects the combined posterior distributions for the stars in each group. The low-$\mathrm{[Mg/Fe]}$ and high-eccentricity stars (A) are slightly  younger than the majority of the high-$\mathrm{[Mg/Fe]}$ sample (C), suggesting that much of the \emph{in-situ}, high-$\mathrm{[Mg/Fe]}$ population was already in place before the major accretion event occurred.}
	
	\label{fig:summary}
\end{figure}


The blue (BS) and red (RS) sequences revealed by Gaia-DR2\cite{Babusiaux2016} in the color-magnitude diagram (CMD) of the kinematically defined halo, have been associated to a population of extra-galactic origin, and to the \emph{in-situ} halo and/or heated-thick disc respectively\cite{Haywood2018}. As shown in Fig.~3, most of the stars of our group (C), which we classified as \emph{in-situ} high-$\mathrm{[\alpha/Fe]}$ disc/halo population, naturally occupy the RS, while likely GES stars lie in the BS.The high precision ages afforded by asteroseismology allow us to order chronologically  the formation of the accreted population with respect to the high-$\mathrm{[\alpha/Fe]}$ \emph{in-situ} one.

Figure~4 displays our main finding: the distribution in age and orbit eccentricity (coloured by $\mathrm{[Fe/H]}$) of stars in our sample.  The  top panel shows the marginalised posterior distributions in age for our three groups of stars: (A) $\mathrm{[Mg/Fe]}$ below the cut and $e>0.7$ (GES debris), all grouped at a similar age and $e$; (B) $\mathrm{[Mg/Fe]}$ below the cut and $e < 0.7$; and (C) $\mathrm{[Mg/Fe]}$ above the cut, with a large spread in $e$, and the oldest ages but with a marked tail of younger stars in the population.

As in other recent papers\cite{Chiappini2015,Martig2015}, we find in our sample a fraction of `apparently young' stars, despite chemical markers indicative of old ages (elevated $\alpha$-element abundances and a high C/N ratio). Since asteroseismology assigns a high mass to these targets, they have previously been identified as `over-massive' $\alpha$-rich stars (likely product of mass transfer\cite{Jofre2016}). 



\begin{table*}[t]
	\label{tab:HBM}
	\centering
	\caption{\small {\bf Properties of the inferred age distribution of the three populations.} Median and $\mathrm{1\sigma}$-error interval of:  mean ages $\mu$, intrinsic age spread $\tau$ of the main population and contaminant ($\mu_c$, $\tau_c$) population (that of  `over-massive' stars), and the contaminant fraction $\epsilon$  for the three populations of stars defined in the sample of metal poor \textit{Kepler} giants. The high- and low-$e$, low-$\mathrm{[Mg/Fe]}$ stars have significantly different age distributions. The high-$e$, low-$\mathrm{[Mg/Fe]}$ stars, which are likely \textit{ex-situ} in origin have a similar (but slightly younger) mean age compared to the majority of the stars in the \emph{in-situ} high-$\mathrm{[Mg/Fe]}$ population. This suggests that these \emph{ex-situ} high eccentricity stars were likely to be formed at roughly the same epoch as, or even after, the high-$\mathrm{[Mg/Fe]}$ population. The contamination by the over-massive (and therefore young in appearance) stars is of the order $10\%$, with a consistent age and spread among each population.}
	\medskip
	\vspace*{5mm}
	\begin{tabular}{lccccc}
		\hline
		Group     &  $\mu\rm[Gyr]$ & $\mathrm{\tau\,[Gyr]}$ & $\mathrm{\mu_{c}\,[Gyr]}$ &  $\mathrm{\tau_c\,[Gyr]}$ & $\epsilon$\\
		\hline
		(A) Low $\mathrm{[Mg/Fe]}$, $e > 0.7$ 
		& $9.7\pm0.6$ 
		& $0.8^{+0.9}_{-0.4}$ 
		& $4.5\pm2.0$
		& $2.9^{+5.7}_{-2.0}$
		& $0.15^{+0.12}_{-0.08}$\\
		
		(B) Low $\mathrm{[Mg/Fe]}$, $e < 0.7$  
		& $8.2\pm0.8$ 
		& $0.8^{+1.0}_{-0.5}$ 
		& $4.9\pm2.0$
		& $2.8^{+5.1}_{-1.8}$
		& $0.06^{+0.07}_{-0.03}$\\
		(C) High $\mathrm{[Mg/Fe]}$ 
		& $10.4\pm0.3$ 
		& $0.5^{+0.4}_{-0.3}$ 
		& $4.2\pm0.8$
		& $2.1^{+4.2}_{-1.4}$
		& $0.16^{+0.05}_{-0.04}$\\
		
		\hline
	\end{tabular}
\end{table*}

We fit a hierarchical model to the stellar ages in each group, assessing the mean age and the intrinsic age spread of each population. We assume that the true age of each star in each group is drawn from a normal distribution with a mean age $\mu$ and width $\tau$, contaminated by a wider normal distribution by some fraction $\epsilon$, with some mean $\mu_c$ and spread $\tau_c$ that captures the contribution of `over-massive' stars (see \textbf{Methods} and Extended Data Fig.~11 for a more detailed description). The best fit parameters for each population are shown in Table~1. These indicate that: \textit{i)} The mean age ($\mu$) of group (B) is significantly lower than that of the population (A) and (C) stars. Although we are aware that some of its elements may be thin disc contaminants, we note that only two of seven have a maximum vertical excursion over the Galactic plane lower than 1~kpc, and removing them from the sample does not change the age distribution.  This difference is visible in the posterior age distribution for these stars shown in Fig.~4 (yellow histogram). \textit{ii)}  Population (A) stars (which we associate with the GES progenitor) have a mean age and spread consistent with those of population (C). This suggests that these stars, which are likely to have been born \emph{ex-situ}, were formed contemporaneously to, if not slightly after, the high-$\mathrm{[\alpha/Fe]}$ population (C) that was formed in the Milky Way starting roughly $\sim 10$ to $11.5$ Gyr ago, as shown by this work. 

The precise ages inferred here for individual objects in the blue and red sequences of Fig.~3, provide crucial constrains to MW formation models and to the more general debate on the dominant drivers of thick disk formation, mergers or cold gas accretion. Some recent  studies\cite{DiMatteo2019, Chaplin2020, Gallart2019, Vincenzo2019, Grand2020, Belokurov2018, Belokurov2019,  Mackereth2018, Myeong2019, Kruijssen2020, Bignone2019} show indirect evidence in favor of a scenario in which the merger with GES may have influenced the evolution of an already existing high-$\mathrm{[\alpha/Fe]}$ proto-Galaxy in some way.  By determining observationally the chronology of events in the early Milky Way with precise ages, our results confirm this emerging picture from other studies which suggests the high-$\mathrm{[\alpha/Fe]}$ population had already formed when GES merged with the Milky Way.

That stars in the Galactic halo belonging to the chemo-kinematically determined accreted GES debris were formed contemporaneously with, or more recently than those of the early \emph{in-situ} MW, has profound implications for the formation and assembly history of the Galaxy. Although the low-metallicity tail of the thick disc ($\mathrm{[Fe/H]<-0.5}$) we have analysed here is expected to be older than its main component, a recent study\cite{Miglio2020} shows that the thick disc population ($\mathrm{-1.55<[Fe/H]<0.26}$ and $\mathrm{[\alpha/Fe]>0.1}$) has an intrinsic age dispersion of only 1.25~Gyr. These results suggest that  a majority of the stars in the high-$\mathrm{[\alpha/Fe]}$ population was in place before the merger with the GES progenitor (which has a lower limit of ~9~Gyr in our age measurements, consistently with the predicted merger time from recent cosmological simulations\cite{Kruijssen2020}), and support models in which GES was \emph{not} a major trigger to the formation of the thick disc. Since it is well established that such $\mathrm{[\alpha/Fe]}$ stars can only form in the most intense star formation events in gas-rich galaxies\cite{Chiappini1997,Mackereth2018}, this implies that either: a) the Galaxy had an extremely gas rich merger \emph{prior} to GES, or b) the early in-situ gas content of the Milky Way was accreted sufficiently fast to form the high-$\mathrm{[\alpha/Fe]}$ stars without any merging event. While the former \emph{predicts} the possible presence of another major merger, the latter suggests that the early dark matter assembly of the Galaxy was rapid enough that it could have accreted gas in sufficient quantities at early times to trigger this star formation.

\newpage

\section*{Methods}

\noindent\textbf{APOGEE, Gaia and \emph{Kepler} data}\\
We select targets with SDSS-IV/APOGEE spectra and NASA-\emph{Kepler} light curves by cross matching the APOGEE DR14 catalogue\cite{apogee,Holtzman2018} with the Kepler Input Catalog (KIC \cite{KIC}). We then cross-match again with the Gaia-DR2 catalogue\cite{gaia}, which provides parallax, position and proper motion measurements for the relevant stars. APOGEE provides, in addition of atmospheric parameters (effective temperature and detailed abundances for 23 different chemical elements), highly precise ($\sim $1\%) radial velocities for all targets. Radial velocities, combined with Gaia proper motions, are used to derive the orbital parameters.

Since we are interested in using asteroseismology to study stellar populations which are likely part of the Milky Way halo, we make a first broad cut to select stars with APOGEE-$\mathrm{[Fe/H] < -0.5}$. We also remove stars with flags from APOGEE which suggest their spectra or the parameters derived from them are unreliable (specifically, we remove stars with \texttt{STAR\_BAD} or \texttt{STAR\_WARN} flags). This leaves a sample of 400 stars with good data from APOGEE and \emph{Kepler}  (see Extended Data Fig.~5) upon which the analysis on the basis of the light curves (described below) can be made. 
\\
\\
\noindent\textbf{Distances}\\
The distance estimates using parallaxes from Gaia-DR2 for our sample have a mean relative error of 15\% (median 11\%), and for 22\% of the sample that value is larger than 20\%.  We also take distance estimates from the \texttt{astroNN} catalogue \cite{LeungBovy2019}, which are based on neural network models of the  APOGEE spectra, trained on the Gaia-DR2 parallaxes.  \texttt{astroNN} distances, which have relative uncertainties of roughly 10\%, and provide a more robust measure of the stellar distances than the parallax information for these more distant stars, have been used in the determination of the orbital parameters.
\\
\\
\noindent \textbf{Orbital parameters}\\
Orbital parameters are estimated for the sample in question using the Stäckel approximation based fast orbit estimation method\cite{MackerethBovy2018} implemented in  \texttt{galpy}\cite{Bovy2015}. We take 100 samples of the covariance matrices for each star, formed from the observed RA, Dec., proper motion in RA and Dec., distance and radial velocity and their uncertainties and correlation coefficients (in this case, the distance and radial velocity are measured independently, so their uncertainties are uncorrelated). We then estimate the orbital parameters for each of these samples assuming the simple \texttt{MWPotential2014} potential, which is adequate in this case, since the majority of these stars have halo-like kinematics and are not likely to be affected by non-axisymmetries in the disc and bulge. We assume the position of the Sun to be $R_0 = 8.125$~kpc (Galactocentric distance, ref.\cite{GRAVITY2018}), and $z_0 = 0.02$~kpc (height above the Galactic mid-plane, ref.\cite{BennettBovy2019}), and its velocity to be $\vec{v}_0=[U,V,W]=[-11.1,245.6,7.25]\ \mathrm{km\ s^{-1}}$  (in the left-handed cartesian Galactic coordinate system), based on the SGR~A* proper motion\cite{GRAVITY2018b} and the solar motion\cite{Schonrich2010}. We estimate pericentre and apocentre radii, orbital eccentricity and the maximum vertical excursion, their uncertainties and correlation coefficients for each star. These orbital parameters will later allow us to verify the accreted nature (or not) of stars in our sample.
\\
\\
\noindent \textbf{Seismic data}\\
We retrieve \emph{Kepler} light curves from MAST ({https://archive.stsci.edu/kepler/publiclightcurves.html}) and measure individual radial-mode frequencies following the approach in ref.\cite{Davies2016}. These results were cross-matching with the radial frequency modes using the automatic pipeline PBJam\\
({https://github.com/grd349/PBjam}), and with the Kallinger's RG-catalogue \cite{Kallinger19} for the targets in common. Although our main results are based on fitting individual-mode frequencies, we have performed additional tests (see below) working with average seismic indexes\cite{Mosser2009,Mosser2011} \deltanu\ and \numax\ (mean large frequency separation and frequency at maximum power, respectively). For a fraction of the stars (90 targets) it was also possible to estimate the value of the gravity-mode period spacing \cite{Vrard2016}.

We measured frequency of at least 3 individual radial modes in 276 targets over 400.  From that sample, we remove the stars in the red clump ($\sim 50\%$). Their current masses (those inferred from seismology) are likely the result of some  mass loss in previous evolutionary phases, and hence their age estimates would depend on the poorly known mass loss prescription itself. This classification is based on the value of the gravity-mode period spacing \cite{Bedding2011,Mosser2014} when available, and from visual inspection of the power spectra \cite{Montalban10}. This classification has been also cross-checked with results in other studies \cite{Yu2018,Kallinger19}.  Among the non core-He burning giants, we restrict the sample to stars with \numax\ larger than 15~$\mu$Hz, this mitigates contamination from early AGB (Asymptotic Giant Branch) stars, and removes stars with relatively low \numax, a domain where seismic constrains are less numerous (the number of radial modes decreases with \numax) and robust.
\\
\\
\noindent \textbf{Final Sample}\\
After the above refereed cuts, our final sample contains 105 stars, likely  in the red giant branch (RGB), with at least four radial modes detected (8,19,47,31 with 4, 5, 6 and 7 modes respectively). Their frequencies have a mean  uncertainty of 0.085\% (median 0.055\%). The typical uncertainty of the effective temperature ($\mathrm{T_{eff}}$) is $\sim 83\,K$, and that of \numax\ is 1.7\%. The characteristic metallicity of the sample is $\mathrm{[Fe/H]}=-0.66$, with 25\% of the targets having an iron content lower than -0.85.  The typical error quoted in  APOGEE-DR14 for $\mathrm{[Fe/H]}$ is $\sim 0.008$. That is substantially smaller than the typical accuracy of  APOGEE-DR14 chemical abundances, as assessed using different and independent spectroscopic analyses of APOGEE stars\cite{Jonsson2018}, and  10 to 20 times smaller the step used in the grid of models. Hence, we increase and  fix the typical error in $\mathrm{[Fe/H]}$ in APOGEE DR14  to $0.05$ dex.

Concerning the $\alpha$ elements, 50\% (40\%, 8\%, and 2\%) of the targets have $\mathrm{[\alpha/Fe]\sim 0.2}$ (0.3, 0.1, and 0.4 respectively).
\\
\\
\noindent\textbf{Bayesian stellar parameter inference}\\
The stellar parameters of each star in our sample have  been estimated by using the open-source code AIMS\cite{aims,Lund_Reese2018,Rendle2019}  (Asteroseismic Inference on a Massive Scale), that implements a Bayesian inference approach. AIMS evaluates the posterior probability distributions of the stellar parameters using a Markov Chain Monte Carlo (MCMC) ensemble sampler\cite{Foreman-Mackey2013}, and selects stellar models that best fit observation data by interpolating (evolutionary tracks and frequencies) in a pre-computed grid. As demonstrated by several works\cite{Lebreton_Goupil2014, Rendle2019,  Miglio2017}, using individual frequencies as observational constrains contributes to significantly reduce the uncertainties affecting estimated global stellar parameters with respect to the precision and accuracy resulting from scaling relations.  The drawback of using individual frequencies is that theoretical values should be corrected by the surface effects\cite{Gough1990}. In this work we use the frequencies of individual radial modes and their uncertainties as observational seismic constraints, and correct the theoretical frequencies using a two-terms prescription\cite{Ball_Gizon2014}. The surface effect corrections involve in that case two free parameters ($a_{-1}$ and $a_{3}$, eq.(4) in ref.\cite{Ball_Gizon2014}) to be derived by the fitting procedure for each target. The other parameters to be determined are the stellar mass, the initial mass fraction of metals, and the stellar age. We do not use specific priors for these parameters, except for $a_{-1}$ and $a_{3}$ if an initial calculation has led to unexpected surface-effect corrections. For that cases we re-run AIMS using uniform priors for these parameters, which domain is estimated from the other successful fits.  As ``classical" constrains we adopt the spectroscopic values of effective temperature and  surface metal content  from APOGEE-DR14, and the average seismic index \numax\ from the analysis of \emph{Kepler} light curves.   Theoretical values of \numax\ cannot be derived from adiabatic oscillation spectra, hence \numax\ for models relies on the scaling relation  $\mathrm{\numax=\numax_{\odot}\cdot\,M(M_\odot)/R^2(R_\odot)/{\sqrt T_{eff}/T_{eff,\odot}}}$ (ref.\cite{Brown91}), with \numax=3090~$\mu$Hz (ref.\cite{Huberetal2011}), and M and R the total stellar mass and radius, respectively, in solar units. To check for consistency, we also run AIMS replacing \numax\ with the bolometric luminosity derived from Gaia-DR2 parallax (see below).

For the purposes of this study, the specific non-solar-scaled chemical composition ($\mathrm{\alpha}$-enhanced) of halo/thick disk stars has been taken into account in the computation of a new grid of stellar models and their adiabatic oscillation frequencies. The stellar models have been computed using the stellar evolution code CLES\cite{cles}, and  following the evolution from the pre-main sequence up to a radius of 25~${\rm R_{\odot}}$ on the RGB phase. The details on the adopted  physics prescriptions and the computed evolutionary tracks are available at {http://doi.org/10.5281/zenodo.4032320}.  The number of models along a track has been chosen  to provide a difference of mean large frequency separation between consecutive models of the order of 0.5\%, and the oscillation frequencies of radial modes have been computed for each stellar model using the adiabatic oscillation code LOSC\cite{losc}.  
\\
\\
\noindent\textbf{Inferred stellar parameters}\\
We get reliable stellar parameters for 95 targets of the 105 classified as RGBs. This final selection is based on the value of the likelihood and on the consistency between parameters inferred using different observational constraints. For instance, the target KIC~7191496 has been removed from the sample because the ages inferred using \numax\ (equal to 16.14 $\mu$Hz) or luminosity differ by more than $5\sigma$. Half of the stars removed from the sample have a value of \numax\ only marginally larger than the threshold  of $15\,\mu$Hz (between 15 and $17.5\,\mu$Hz). For KIC~3630240, which has a high value of \numax\ (close to the Nyquist frequency for these time series), AIMS does not converge if luminosity is used as observational constrain (regardless of the adopted Gaia zero-offset parallax).

The properties of the 95 stars of our final sample are collected in the Supplementary Table~1. Its last columns contains the values of stellar mass, radius and age of the model that best match observations, as well as their $\pm\mathrm{1\,\sigma}$ interval values. These are internal statistical errors based on the sampling of the posterior probability distributions obtained with the grid of models used. Extended Data Figs.~6,~7 present the posterior distributions for six relevant stellar parameters for the targets KIC~4143467 and KIC~12111110 respectively. In both cases these distributions were obtained using as constraints in AIMS:  6 radial modes, surface mass fraction of metals, effective temperature and \numax.  KIC~4143467 is one of the targets shown in Fig.~1, while KIC~12111110 is the object at  age $\sim10$~Gyr, and  eccentricity 0.99 (Fig.~4) which shows a large and very asymmetric uncertainty.  In this case the posterior distributions are clearly bi-modal. Although the best match with observation (dot-dashed vertical line) is achieved for the older group of models,   a large number of young models still have a high probability.  An uncertainty in luminosity smaller than 10\% should be needed to critically discriminate between the two solutions. 
\\
\\
\noindent\textbf{Robustness tests against systematic uncertainties}\\
It is widely accepted that the effective temperature of RGB models strongly depends on the convection mixing-lenght parameter ($\alpha_{\rm MLT}$) and on the adopted atmosphere boundary conditions. A systematic difference between ${\mathrm{T_{eff}}}$ of the models and observations could indicate that those parameters are not the adequate ones to represent observational data, creating  a tension, leading to systematic larger or smaller stellar masses, and hence affecting the estimated ages. To check for the robustness of our stellar dating, we have run AIMS for all the stars after shifting their effective temperature by $\mathrm{\pm\,85\,K}$ (AIMS results -with the grid of models above described- are typically 85~K hotter than observed  $\mathrm{T_{eff}}$). Although the likelihood is generally higher for the temperature scale shifted by $\mathrm{+85\,K}$, the stellar parameters retrieved do not change.
In fact, the fitting is dominated by the individual frequencies, with a lower impact of $\mathrm{T_{eff}}$, directly or trough \numax.
\\
\\
\noindent
In the fitting process we do not interpolate in the parameter ${\rm [\alpha/Fe]}$, but we select the grid with the closest value to that estimated from spectroscopy. In order to estimate the effect of the ${\rm [\alpha/Fe]}$  step on the derived stellar parameters, we have run AIMS using grid of models computed with ${\rm [\alpha/Fe]}$ values shifted by $\pm0.1$~dex, and compared their ages. The  differences between ages obtained using the nominal ${\rm [\alpha/Fe]}$ or that shifted, divided by their uncertainties,  have a standard deviation of 0.3.
\\
\\
\noindent
AIMS allows different prescriptions for the surface effects correction. We check the effect of using the one term\cite{Ball_Gizon2014} prescription. The differences between one and two terms prescriptions depends on the number of modes observed. For 4 modes, the results with the two approaches are in good agreement. However,  as the number of modes increases, one-term prescription appears clearly inefficient fitting observed oscillation frequencies, and  systematically provides much younger ages.
\\
\\
\noindent
We expect that a large part of the AGB contamination will have been removed from the sample by filtering out \numax\ values smaller than 15~$\mu$Hz and selecting targets with at least 4 radial modes. Nevertheless, we test the effect on the age distribution of a miss-classification RGB/AGB. We select synthetic AGB models and spectra, and derive their stellar parameters using AIMS, as above described, that is, with a grid of models which stop at $\mathrm{R=25\,R_\odot}$ in the RGB phase.  As a consequence of the miss-classification AIMS either,  does not converge, or provide very high (and unrealistic - $20-25$~Gyr) stellar ages. 

Computing $\ell=2$ oscillation modes is very time consuming and we have not used then in the fitting process. However, their frequencies for our sample have been determined at the same time than the radial ones, allowing us to estimate the mean value of the small frequency separation ($\delta\nu_{02}=\nu_{\ell=0,n}-\nu_{\ell=2,n-1}$). This seismic index, which values depend also on the evolutionary state,  is a good proxy of the stellar mass\cite{Montalban10,Montalban12,Handberg15} and can help us to identify genuine massive stars. The trend of  $\delta\nu_{02}/\Delta\nu_0$ is generally consistent with the masses assigned by AIMS assuming that our sample is formed by RGB. For a subset of 22 targets we also have measures of the dipole-mode period spacing. These values and the $\delta\nu_{02}$ ones are consistent with classifying these targets as RGB stars, including among them  two of the massive/young stars. 
\\
\\
\noindent\textbf{ Cross-checking using Gaia parallaxes}\\
We check the consistency of derived stellar parameters with luminosity values from Gaia-DR2. These values are obtained using Gaia-DR2 parallaxes, 2MASS $K\mathrm{_s}$ apparent magnitudes, and bolometric corrections\cite{Casagrande2018}  appropriate to the atmosphere parameters (seismic surface gravity, and spectroscopic $T_{\rm eff}$ and photospheric chemical composition).  The most important contribution to the luminosity uncertainty comes from
parallax, which suffers from a zero-point offset of the order of few tens $\mu as$\cite{Lindegren2018,Khan2019, Hall2019,Zinn2019}. The effect of different extinction estimates\cite{Green2015,Green2017,Green2019,Rodrigues2015} turns out to be only 0.008~dex for the magnitude $K\mathrm{_s}$.

We perform two new runs of stellar parameter estimation using luminosity (with offset of 30~$\mu as$ and 50~$\mu as$) instead of \numax\ as observational constraint in AIMS. The results are generally in good agreement with previous ones. The fits are dominated by highly precise frequencies and the still large errors affecting luminosity do not allow us to discriminate in case of multi-modal posterior distribution  nor to check the reliability of \numax\ scaling law.

The stellar radii derived from the Stefan-Boltzmann law with the spectroscopic effective temperatures and Gaia luminosity values have been compared with those inferred using AIMS. The residuals divided by the relevant uncertainties have a standard deviation close to one, with no apparent trend with e.g. \numax. The median offset is, on the other hand, sensitive to the assumed zero-point parallax offset (better agreement with 50~$\mu as$ zero-point parallax offset) and to the effective temperature scale (consistently with results above). 
\\
\\
\noindent\textbf{Cross-checking using global seismic parameters for a larger set of  \emph{ex-situ} and \emph{in-situ} stars}\\
As an additional test for robustness we use the code PARAM\cite{Rodrigues2017}
to infer masses, radii and ages for the larger set of stars (Extended Data Fig.~5).  We consider all stars (RC and RGB) with average seismic parameters ($\mathrm{\Delta\nu}$, \numax) determined by the COR pipeline\cite{Mosser2011}.
While the results from the detailed analysis are more precise and more accurate (the median age uncertainty given by PARAM is 25\% instead 11\%), we use the age distribution of the wider sample to check if our main results are compatible with those from an independent modelling code and using average seismic parameters. The grid of models (computed with the code MESA\cite{Paxton2013}) at the base of these calculations includes:  a mass loss during RGB evolutionary phase equivalent to a Reimers' parameter $\eta=0.2$, and a model-based correction for the large frequency separation-mean density relation\cite{Rodrigues2017}. 

The results for two samples selected by limiting the radius to 14 or 8 $\mathrm{R_\odot}$ (including or not the RC, Extended Data Figs.~8,~9) show that the age distributions of the accreted and \emph{in-situ} stars show the same trend as the results obtained by AIMS with a smaller, high quality sample: accreted stars are contemporaneous or slightly younger than \emph{in-situ} ones. 
\\
\\
\noindent\textbf{Modelling the age distributions of stellar groups defined on the basis of element abundances and kinematics}\\
We model the intrinsic age distribution of the populations in $\mathrm{[Fe/H]}$-$\mathrm{[Mg/Fe]}$ and kinematics space selected in Fig.~4 using a simple hierarchical Bayesian model (HBM, see graphical model in Extended Data Fig.~11). We expect that the stars in these groups are likely to belong to \emph{in-} and \emph{ex-situ} stellar populations, and therefore such a modelling provides a means of statistically comparing the age distributions whilst taking the age uncertainties properly into account. We assume that age measurements of stars in a given population are drawn from a normal distribution with a mean age $\mu$ and intrinsic age spread $\tau$, with some measurement error $\sigma_{\mathrm{age}}$ (derived from the posterior probability given by AIMS). We include an outlier term in our model, assuming that in each population there is an over-density at younger age due to our measurement of stars which are `over-massive' (likely due to binary interactions) and therefore appear young. We assume that these outliers are also distributed normally with a mean age $\mu_c$, a spread $\tau_c$ and contributing some fraction $\epsilon$.

While  we have no good way of determining whether these stars are bona-fide young stars or indeed, over-massive stars, we are confident that this would not fundamentally change the results of the HBM analysis we perform. In particular, since the model always assumes contamination at younger ages, the only effect of including some bona-fide young stars (which, as we have determined, are likely separated from the GES stars) would be to artificially increase the value of our epsilon (contamination fraction) parameter, and would not significantly alter the mean-age of the target population.

We sample the posterior probability distribution given the data in each group in element abundance space using \texttt{pymc3}. We make use of the the No-U-Turn-Sampler (NUTS), a variant of Hamiltonian Monte Carlo, which uses the gradients of the likelihood function to facilitate rapid convergence and sampling of the posteriors over many parameters. For each population, we take 1000 samples of the posterior over 4 independent chains after allowing 1000 burn-in steps, for a total of 4000 samples. 
\\
\\
\noindent\textbf{Effect of selection criteria}\\
The classification in \emph{in-} and \emph{ex-situ} populations of our low-metal sample ($\mathrm{[Fe/H]<-0.5}$) is based on a particular cut in the $\mathrm{[\alpha/Fe]}$ plane  and in eccentricity. Unfortunately, there is no consensus in literature on which cut should be adopted nor on the value of $\mathrm{[Fe/H]}$ defining the low-metallicity end of the Galactic thin disc. 
We have analysed the effect of considering different division lines in the $\mathrm{[\alpha/Fe]}$ plane and of shifting the threshold of eccentricity from 0.7 to 0.6 (Extended Data Fig.~10). Besides the selection made in the main paper,  we have used two other division lines: $\mathrm{[Mg/Fe]=-0.5\,[Fe/H]-0.3}$ (used in ref.\cite{Mackereth2019}), and $\mathrm{[Mg/Fe]=-0.2\,[Fe/H]}$\cite{Hayes2018,Fernandez-Alvar2018} (which has the same slope than the one used in the main paper but with a zero-point of 0 instead of 0.05). The conclusions of the paper do not change: the \emph{in-situ} high-$\alpha$ population (group C) is slightly older than that of GES (group A).   Different selection criteria modify mainly  the composition of the group (B), which is not associated with a particular stellar population. It contains  what is not in (A) or (C), and in particular may contain some contamination from the low-metallicity end of the thin disc. The two stars at the high-metallicity end ([Fe/H]$>$ -0.7) of out group (B) are likely part of the thin disc as indicated by their orbital and chemical properties. They can end in a different group depending on the cut used, however, that does not fundamentally change the age distributions of the other two groups. We believe that the contamination from the thin disc does not affect other stars of group (B), since these two are the only ones having  a maximum vertical excursion lower than 1~kpc.



\begin{figure}
	\centering
	\includegraphics[width=140mm]{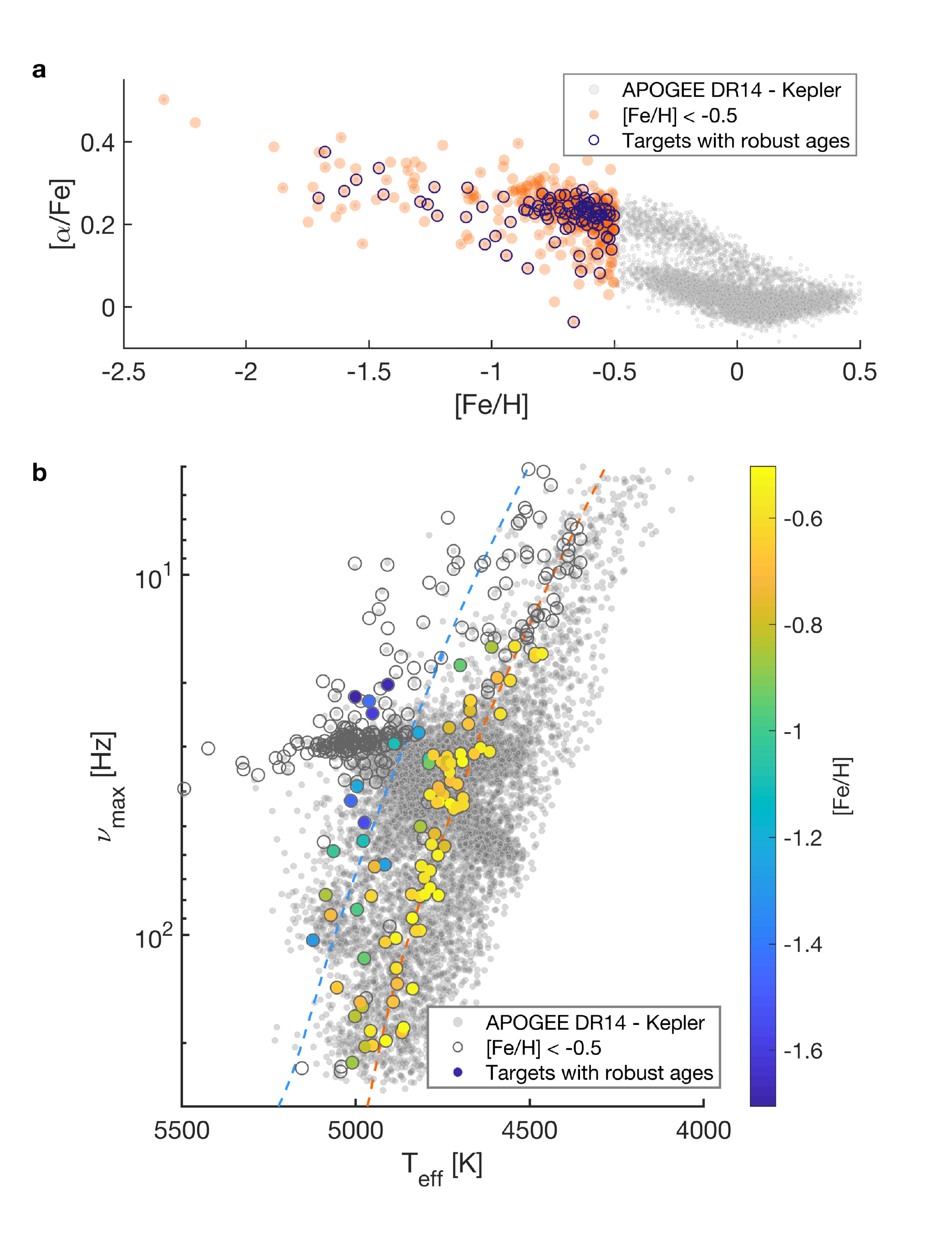}
	
	\caption{\small \textbf{Data samples.} \textbf{a}, Diagram $\mathrm{[\alpha/Fe]}$ versus $\mathrm{[Fe/H]}$ for all the $\emph{Kepler}$-APOGEE-DR14 sample (grey dots). Orange symbols are the targets in our sub-sample: red giant stars with $\mathrm{[Fe/H]<-0.5}$, and blue ones are the first ascending red giant branch targets selected for characterization in this paper.
		\textbf{b,}
		$\mathrm{T_{eff}}$ versus $\mathrm{\nu_{max}}$ diagram (equivalent to Kiel diagram) of our target sample (color-coded by metallicity), overlying  the complete -\emph{Kepler}-APOGEE-DR14 one (grey empty and full symbols). The dashed lines corresponds to two [$\alpha$/Fe]=0.2 evolutionary tracks: blue M=0.9~M$_\odot$ , [Fe/H]=-1.0; orange, same mass but [Fe/H]=-0.5.}
	
	\label{fig:ext1}
\end{figure}
\begin{figure}
	\centering
	\includegraphics[width=165mm]{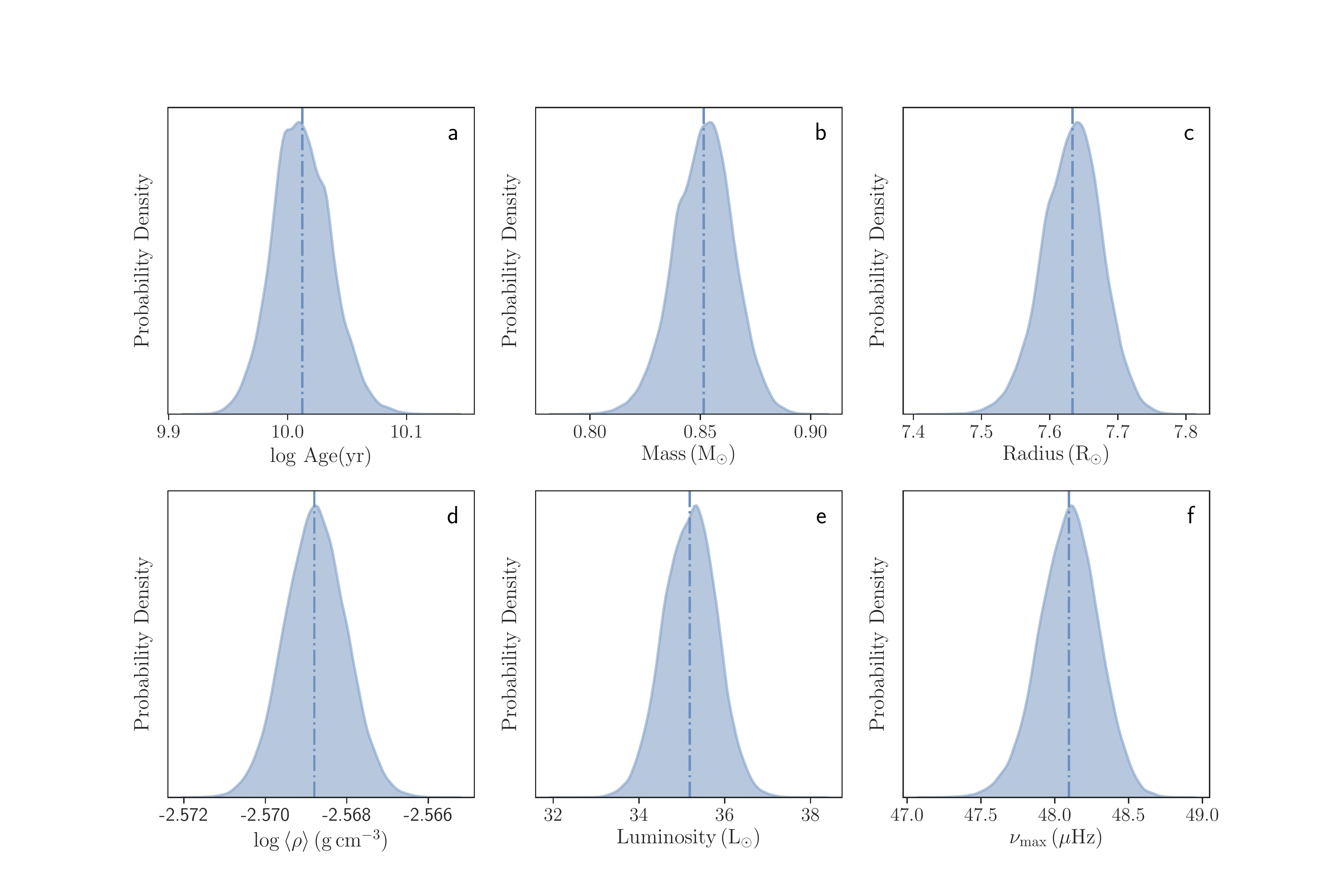}
	
	\caption{\small \textbf{Posterior probability distributions for KIC~4143467 stellar properties as inferred by AIMS.} \textbf{a-f}: age, mass, radius, mean density, luminosity and frequency at maximun power, respectively. The oscillation spectra of this target is shown in first panel of Fig.~1. The vertical dash-dotted lines indicate the value of the corresponding parameter in the best-fitting model from the MCMC sampling.}
	
	\label{fig:ext2}
\end{figure}
\begin{figure}
	\centering
	\includegraphics[width=165mm]{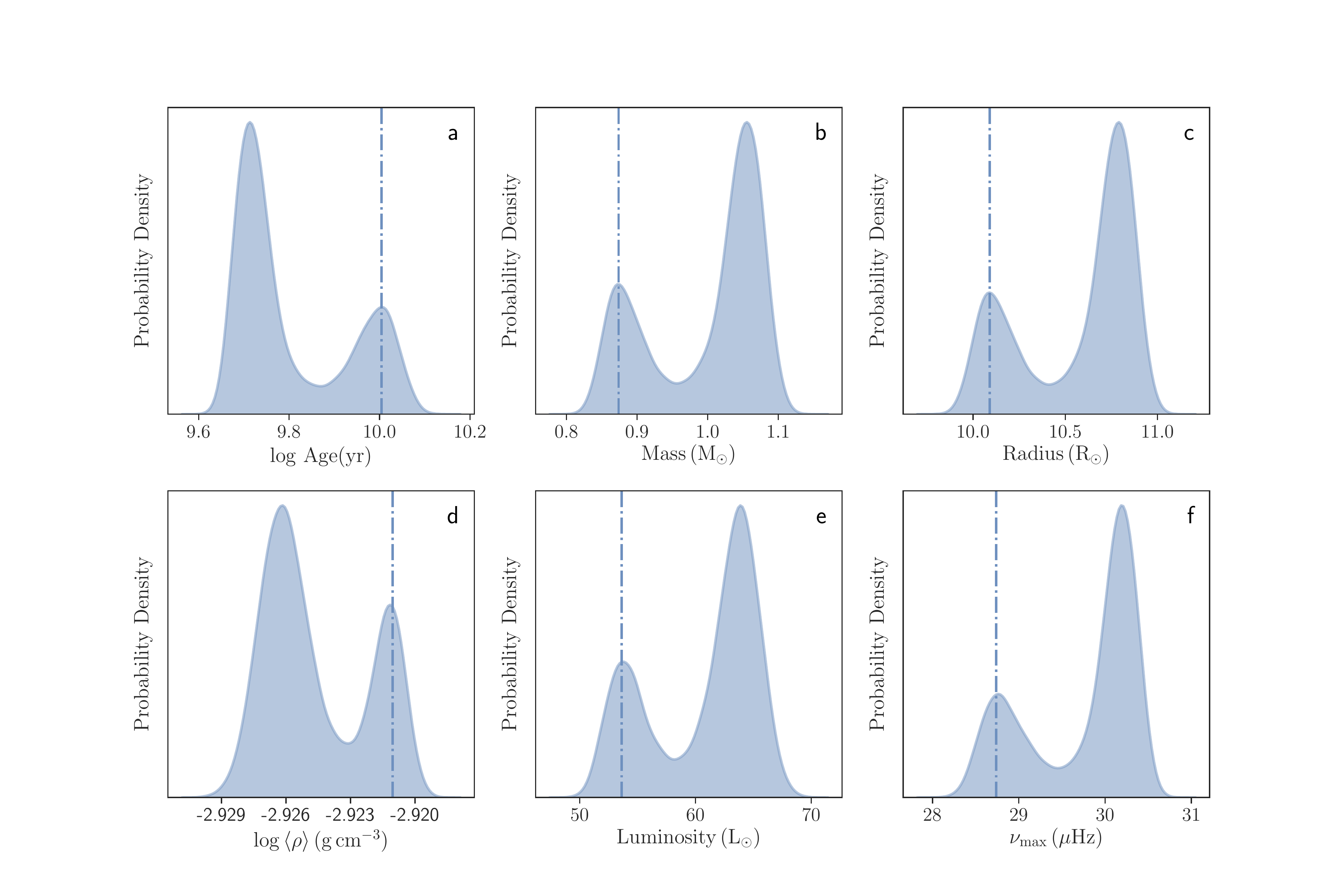}
	
	\caption{\small  \textbf{Posterior probability distributions for KIC~12111110 stellar properties as inferred by AIMS.} \textbf{a-f}: age, mass, radius, mean density, luminosity and frequency at maximun power, respectively. The vertical dash-dotted lines indicate the value of the corresponding parameter in the best-fitting model from the MCMC sampling.}
	
	\label{fig:ext3}
\end{figure}
\begin{figure}
	\centering
	\includegraphics[width=165mm]{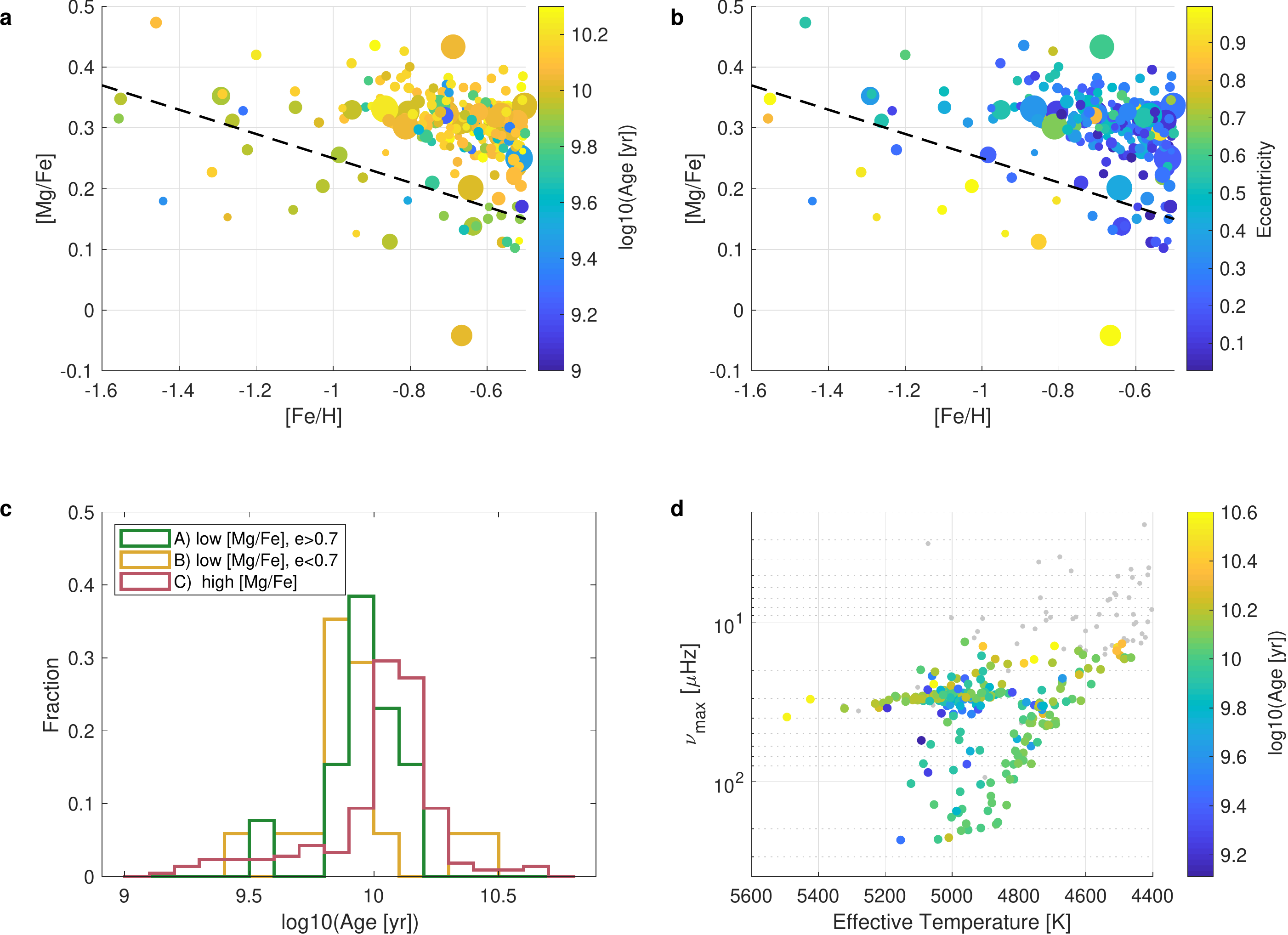}
	
	\caption{\small \textbf{Age distribution using PARAM for the APOGEE-Kepler sample with stellar radius limited to 14~$R_\odot$}. Upper panels: $\mathrm{[\alpha/Fe]}$ vs. $\mathrm{[Fe/H]}$ distribution of the sample coloured by age (\textbf{a}) and eccentricity (\textbf{b}). The symbol size scales with \numax. \textbf{c} panel: Age distributions of accreted and \emph{in-situ} stars,  so classified from their $\mathrm{[\alpha/Fe]}$  and eccentricity values; \textbf{d} panel: Kiel diagram of the sample colored by metallicity (right). Notice that the  ``very old" (yellow dots $\mathrm{T_{eff} > 5400~K}$) suggest that we have underestimated the mass loss for those stars.}
	
	\label{fig:ext4}
\end{figure}
\begin{figure}
	\centering
	\includegraphics[width=165mm]{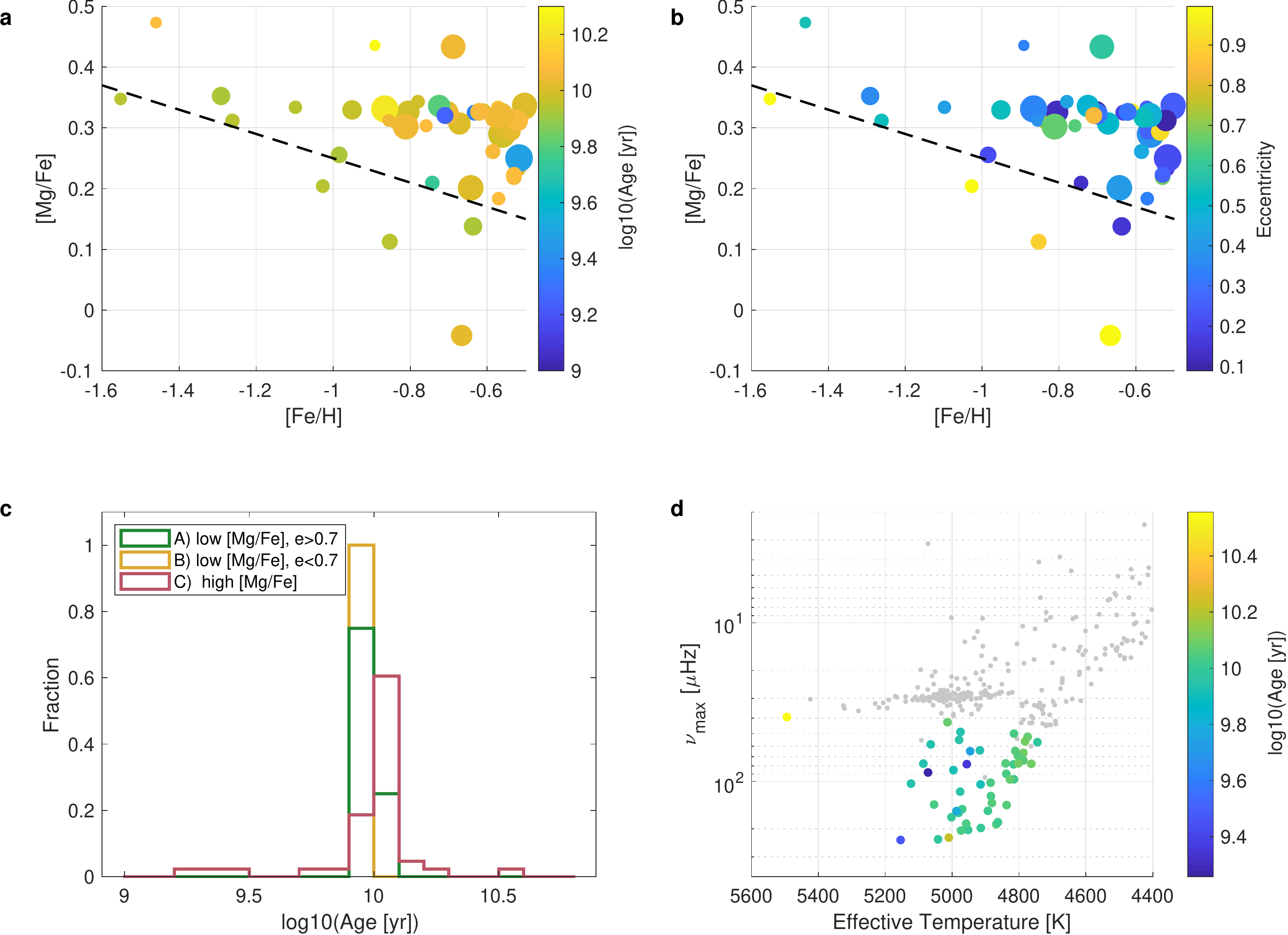}
	
	\caption{\small \textbf{Age distribution using PARAM for the APOGEE-Kepler sample with stellar radius limited to 8~$R_\odot$}. Upper panels: $\mathrm{[\alpha/Fe]}$ vs. $\mathrm{[Fe/H]}$ distribution of the sample coloured by age (\textbf{a}) and eccentricity (\textbf{b}). The symbol size scales with \numax. \textbf{c} panel: Age distributions of accreted and \emph{in-situ} stars,  so classified from their $\mathrm{[\alpha/Fe]}$  and eccentricity values; \textbf{d} panel: Kiel diagram of the sample colored by metallicity (right). Notice that the  ``very old" (yellow dots $\mathrm{T_{eff} > 5400~K}$) suggest that we have underestimated the mass loss for those stars.}
	
	\label{fig:ext5}
\end{figure}


\begin{figure}
	\centering
	\includegraphics[width=130mm]{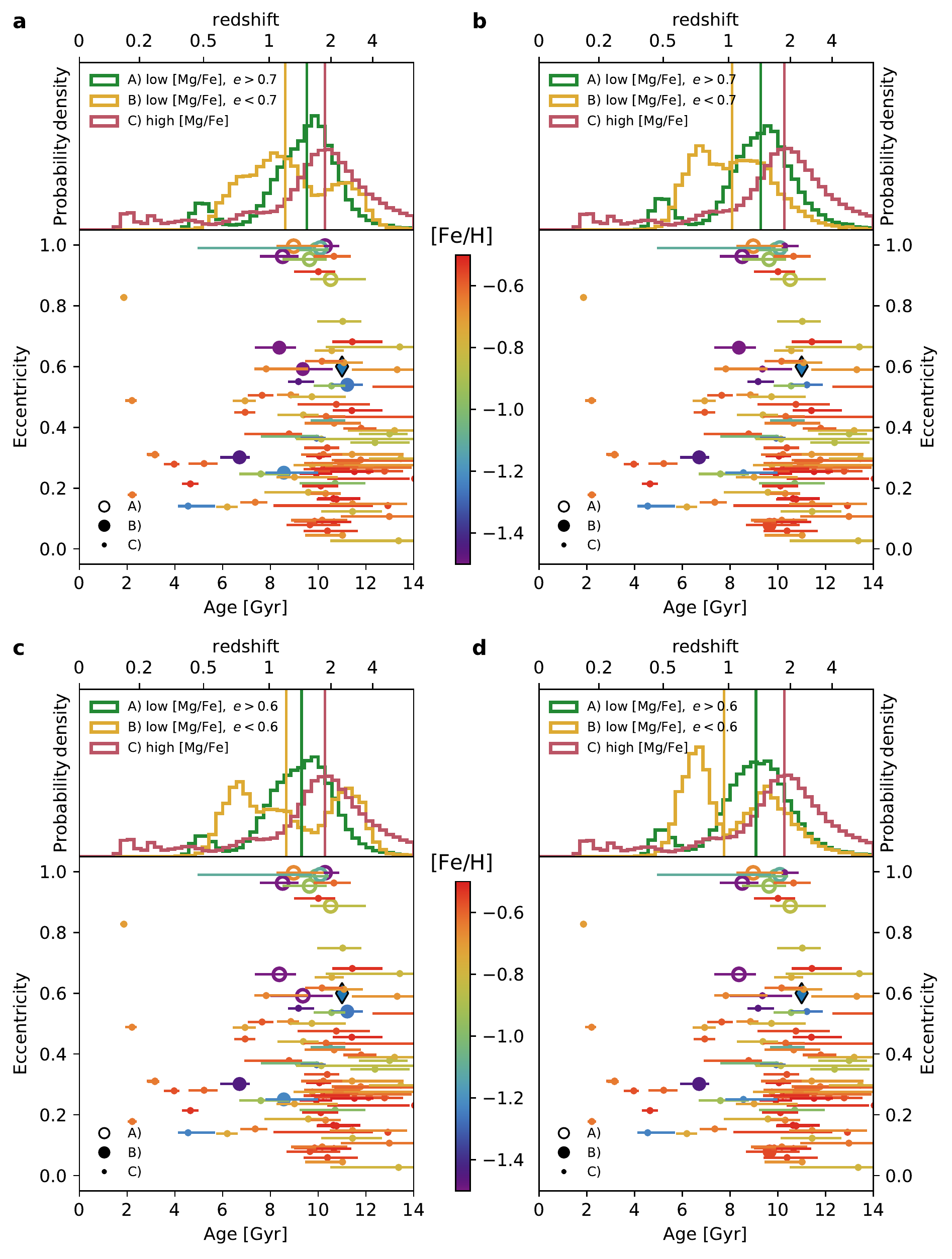}
	
	\caption{\small {\bf Age and eccentricity distributions for different selection criteria for \textit{in-situ} and accreted populations.}  Age against eccentricity $e$ for the stars in the sample coloured by $\mathrm{[Fe/H]}$. Circles respresent age values of the best fitting models, and horizontal lines their uncertainties ([16\%-84\%] C.I. from full posterior distributiosn). Uncertainties on $e$ are smaller than the symbol size. The diamond represents $\nu$~Indi\cite{Chaplin2020} (not included in the distributions). The histogram above reflects the combined posterior distributions for the stars in each selection. \textbf{a, c} panels: division line $\mathrm{[Mg/Fe]=-0.5\,[Fe/H]-0.3}$\cite{Mackereth2019}. \textbf{b, d} panels:  division line $\mathrm{[Mg/Fe]=-0.2\,[Fe/H]}$\cite{Hayes2018}. Top and bottom panels correspond to eccentricity threshold 0.7 and 0.6 respectively.}
	
	\label{fig:ext6}
\end{figure}

\begin{figure}
	\centering
	\includegraphics[width=80mm]{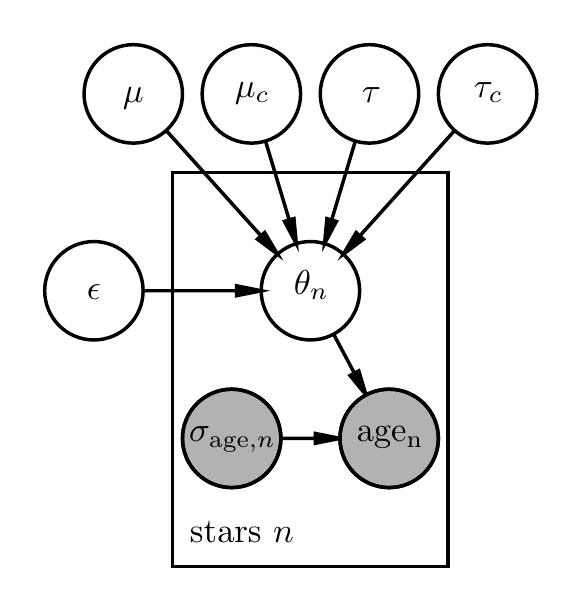}
	
	\caption{\small \textbf{Probabilistic graphical model of that used to fit the mean age and intrinsic age spread of the \emph{in-} and \emph{ex-situ} populations defined on the basis of element abundances and orbital parameters.} We assume the measured ages are drawn from an underlying true age $\theta$ distribution that is Gaussian with a mean $\mu$ with a standard deviation $\tau$. We assume that the true age distribution is contaminated by stars whose mass is higher than expected (and therefore appear younger), likely due to some poorly understood process such as binary interactions. We model these contaminants as also being drawn from another normal distribution with a mean $\mu_c$ and spread $\tau_c$ which has a fractional contribution $\epsilon$ to the total age distribution (hence the main population contributes $1-\epsilon$).}
	
	\label{fig:ext7}
\end{figure}


\newpage


\noindent \textbf{Acknowledgements} JM, JTM, AM, FV, and EW acknowledge support from the ERC Consolidator Grant funding scheme (project ASTEROCHRONOMETRY, G.A. n. 772293). FV acknowledges the support of a Fellowship from the Center for Cosmology and AstroParticle Physics at The Ohio State University. MV is supported by FEDER - Fundo Europeu de Desenvolvimento Regional through COMPETE2020 - Programa Operacional Competitividade e Internacionalização by these grants: PTDC/FIS-AST/30389/2017 \& POCI-01-0145-FEDER-030389.CC acknowledges partial support from DFG Grant CH1188/2-1 and from the ChETEC COST Action (CA16117), supported by COST (European Cooperation in Science and Technology).G.B. acknowledges fundings from the SNF AMBIZIONE grant No 185805 (Seismic inversions and modelling of transport processes in stars) and from the European Research Council (ERC) under the European Union's Horizon 2020 research and innovation programme (grant agreement No 833925, project STAREX). G.R.D. acknowledges funding from the European Research Council (ERC) under the European Union’s Horizon 2020 research and innovation programme (CartographY GA 804752). MBN acknowledges support from the UK Space Agency. OJH acknowledges the support of the UK Science and Technology Facilities Council (STFC).
This article made use of AIMS, a software for fitting stellar pulsation data, developed in the context of the SPACEINN network, funded by the European Commission’s Seventh Framework Programme.
This work has made use of data from the European Space Agency (ESA) mission
{\it Gaia} (https://www.cosmos.esa.int/gaia), processed by the {\it Gaia}
Data Processing and Analysis Consortium (DPAC, {https://www.cosmos.esa.int/web/gaia/dpac/consortium}). Funding for the DPAC
has been provided by national institutions, in particular the institutions
participating in the {\it Gaia} Multilateral Agreement. 
The computations described in this paper were performed using the University of Birmingham's BlueBEAR HPC service, which provides a High Performance Computing service to the University's research community. See http://www.birmingham.ac.uk/bear for more details. The authors thank Professor Sean McGee  for reading and commenting the manuscript. \\


\noindent \textbf{Author contributions} J.M. led the project, with the help from J.T.M., A.M., F.V., C.C. and W.J.C. G.R.D. designed the pipeline for the light-curve analysis. J.M., J.T.M., A.M., F.V., G.R.D., O.H., M.B.N., S.K. and W.E,vR. worked on extracting mode parameters from the \emph{Kepler} light-curves. J.M., A.N. and R.S. performed the stellar modeling and theoretical oscillation frequency computations. J.M., G.B. and B.M.R.  worked on the stellar parameter determination from individual frequencies using  Bayesian inference code AIMS. A.M. estimated stellar parameters from global observational constraints using the code PARAM. B.M. and M.V. provided global seismic parameters. J.T.M. and F.V. performed the kinematics  and  chemical composition analysis  from Gaia-DR2 and APOGEE-DR14 data sets. E.W. derived absolute stellar luminosity from Gaia-DR2. J.W.F. provided radiative opacity data at low temperature for the alpha-enhanced chemical mixture used in the stellar evolution code. All authors have contributed to the interpretation of the data and the results, and discussion and giving comments on the paper.\\


\noindent \textbf{Competing Interests} The authors declare that they have no competing financial interests.\\


\noindent \textbf{Correspondence}  Correspondence and requests for materials should be addressed to J.M. (email: josefina.montalban7@gmail.com).\\


\noindent \textbf{Data Availability}\\
All raw observational data are publicly available: \emph{Kepler} light-curves at \\
{https://archive.stsci.edu/kepler/publiclightcurves.html}; Gaia-DR2 at {https://gea.esac.esa.int/archive} and APOGEE Data Release 14 may be accessed via {https://www.sdss.org/dr14/}.  APOGEE-DR14 raw data have been used in top panels of Fig.~2. Processed data such as individual frequencies, orbital and stellar parameters are available in the supplementary material table or on request. Evolutionary tracks are publicly available at {http://doi.org/10.5281/zenodo.4032320} and theoretical stellar models and oscillation frequencies are  available on request. \\


\noindent \textbf{Code Availability}\\
The asteroseismic modelling results were provided by the code AIMS,  available at\\
{https://lesia.obspm.fr/perso/daniel-reese/spaceinn/aims/version1.3} and cross-checked using the code PARAM
({http://stev.oapd.inaf.it/cgi-bin/param}).\\
The peak-bagging was performed using the pipeline PBJam ({https://github.com/grd349/PBjam})\\
and the orbital parameter determination with \texttt{galpy} ({https://www.galpy.org}). The codes CLES and LOSC used for stellar evolution and adiabatic oscillation computations are not publicly available but evolutionary tracks, structure models and oscillation files are available on reasonable request.\\


\section*{Supplementary materials}

\noindent\textbf{Supplementary Table~1: Stellar parameters for the final sample of 95 targets.} First column contains \emph{Kepler}-ID, second one the selection group, and the following seven columns contain some observational constrains (mean value and $\mathrm{1\sigma}$ uncertainty, when relevant): $\mathrm{n}$,  number of radial modes detected; $\mathrm{\nu_{max}}$ from light curve analysis; Large frequency separation from linear fitting of individual frequencies; $\mathrm{T_{eff}}$,  and chemical properties from APOGEE-DR14; and orbital eccentricity, respectively. The last nine columns have the best fit values of mass, radius and ages inferred by AIMS ($\mathrm{X_{BF}}$) and the values corresponding to the 16\% and 84\% of their posterior distributions ($\mathrm{X_{16}}$, $\mathrm{X_{84}}$ respectively).

\import{.}{Supplementary_Table1.tex}

\end{document}

%% file: Supplementary_Table1.tex
\begin{landscape}
\begingroup

\noindent\textbf{Supplementary Table~1: Stellar parameters for the final sample of 95 targets.} First column contains \emph{Kepler}-ID, second one the selection group, and the following seven columns contain some observational constrains (mean value and $\mathrm{1\sigma}$ uncertainty, when relevant): $\mathrm{n}$,  number of radial modes detected; $\mathrm{\nu_{max}}$ from light curve analysis; Large frequency separation from linear fitting of individual frequencies; $\mathrm{T_{eff}}$,  and chemical properties from APOGEE-DR14; and orbital eccentricity, respectively. The last nine columns have the best fit values of mass, radius and ages inferred by AIMS ($\mathrm{X_{BF}}$) and the values corresponding to the 16\% and 84\% of their posterior distributions ($\mathrm{X_{16}}$, $\mathrm{X_{84}}$ respectively).
\\
\\
\begin{longtable}{|c|c|c|r|r|l|c|c|r|r|r|r|r|r|r|r|r|r|}  
\multicolumn{1}{|c|}{KIC} &
\multicolumn{1}{c|}{group}& 
\multicolumn{1}{c|}{n}&
\multicolumn{1}{c|}{$\mathrm{\nu_{max}}$} &
\multicolumn{1}{c|}{$\mathrm{\Delta \nu _{fit}}$} & 
\multicolumn{1}{c|}{$\mathrm{T_{eff}}$} &
\multicolumn{1}{c|}{$\mathrm{[Fe/H]}$} & 
\multicolumn{1}{c|}{$\mathrm{[\alpha/Fe]}$} & 
\multicolumn{1}{c|}{$\mathrm{ecc}$} &
\multicolumn{3}{c|}{$\mathrm{Mass \, (M_\odot)}$} &   \multicolumn{3}{c|}{$\mathrm{Radius \, (R_\odot)}$} & \multicolumn{3}{c|}{$\mathrm{Age \, (Gyr)}$}  \\
\cline{10-12} \cline{13-15} \cline{16-18}
& & & & & & & & & 
\multicolumn{1}{c|}{$\mathrm{M_{16}}$} & 
\multicolumn{1}{c|}{$\mathrm{M_{BF}}$} &
\multicolumn{1}{c|}{$\mathrm{M_{84}}$} &
\multicolumn{1}{c|}{$\mathrm{R_{16}}$} & 
\multicolumn{1}{c|}{$\mathrm{R_{BF}}$} &
\multicolumn{1}{c|}{$\mathrm{R_{84}}$} &
\multicolumn{1}{c|}{$\mathrm{\tau_{16}}$} &
\multicolumn{1}{c|}{$\mathrm{\tau_{BF}}$} &
\multicolumn{1}{c|}{$\mathrm{\tau_{84}}$} \\
\hline
& & & & & & & & & & & & & & & & & 
\endhead
2443903 & C & 6 &  66.21 $\pm$   1.13 &  6.930 &   4787 $\pm$  79 & -0.57 &  0.30 & 0.256 &  0.88 &  0.89 &  0.92 &  6.69 &  6.72 &  6.80 & 11.37 & 12.78 & 13.57     \\
2971380 & A & 6 &  77.37 $\pm$   1.32 &  7.839 &   5086 $\pm$ 110 & -0.85 &  0.17 & 0.887 &  0.85 &  0.89 &  0.91 &  6.10 &  6.19 &  6.24 &  9.66 & 10.52 & 12.02     \\
3553435 & C & 6 &  41.73 $\pm$   0.71 &  4.737 &   4691 $\pm$  91 & -0.63 &  0.30 & 0.293 &  0.88 &  0.91 &  0.93 &  8.49 &  8.60 &  8.68 & 10.58 & 11.78 & 13.46     \\
3936507 & C & 6 &  33.02 $\pm$   0.56 &  4.044 &   4693 $\pm$  85 & -0.50 &  0.17 & 0.165 &  0.91 &  0.94 &  0.96 &  9.70 &  9.80 &  9.88 &  9.97 & 10.64 & 11.75     \\
4136835 & C & 7 &  77.99 $\pm$   1.33 &  7.158 &   4956 $\pm$  84 & -0.63 &  0.29 & 0.178 &  1.42 &  1.46 &  1.49 &  7.81 &  7.87 &  7.93 &  2.03 &  2.21 &  2.40     \\
4143467 & A & 6 &  48.81 $\pm$   0.83 &  5.605 &   4975 $\pm$  88 & -1.55 &  0.34 & 0.997 &  0.84 &  0.85 &  0.86 &  7.59 &  7.63 &  7.67 &  9.75 & 10.29 & 10.87     \\
4345370 & C & 5 &  32.48 $\pm$   0.55 &  4.033 &   4791 $\pm$  77 & -1.04 &  0.21 & 0.370 &  0.89 &  0.91 &  0.96 &  9.58 &  9.67 &  9.85 &  7.58 &  9.15 & 10.02     \\
4637699 & C & 6 &  54.79 $\pm$   0.93 &  6.046 &   4979 $\pm$  96 & -1.10 &  0.26 & 0.422 &  0.86 &  0.87 &  0.89 &  7.26 &  7.30 &  7.35 &  9.68 & 10.48 & 11.13     \\
4664515 & C & 6 & 186.77 $\pm$   3.18 & 15.650 &   4868 $\pm$ 100 & -0.69 &  0.27 & 0.590 &  0.83 &  0.86 &  0.90 &  3.87 &  3.92 &  3.98 & 11.41 & 13.31 & 15.09     \\
4729920 & C & 7 & 153.75 $\pm$   2.61 & 13.326 &   4893 $\pm$  97 & -0.67 &  0.22 & 0.507 &  0.95 &  0.96 &  0.98 &  4.50 &  4.51 &  4.54 &  8.26 &  8.86 &  9.20     \\
4737942 & C & 5 &  24.38 $\pm$   0.41 &  3.110 &   4584 $\pm$  67 & -0.59 &  0.24 & 0.378 &  0.97 &  0.99 &  1.05 & 11.68 & 11.75 & 12.02 &  6.90 &  8.78 &  9.35     \\
4856592 & C & 7 & 202.93 $\pm$   3.45 & 16.596 &   4952 $\pm$  84 & -0.64 &  0.12 & 0.414 &  0.88 &  0.91 &  0.94 &  3.81 &  3.84 &  3.89 &  9.43 & 10.67 & 11.79     \\
5107950 & C & 6 &  43.99 $\pm$   0.75 &  4.933 &   4716 $\pm$  76 & -0.58 &  0.24 & 0.091 &  0.94 &  0.96 &  0.98 &  8.54 &  8.59 &  8.66 &  8.97 &  9.85 & 10.57     \\
5206349 & B & 7 & 104.71 $\pm$   1.78 &  9.765 &   4915 $\pm$  88 & -0.64 &  0.11 & 0.153 &  0.99 &  1.00 &  1.03 &  5.57 &  5.60 &  5.65 &  6.73 &  7.36 &  7.87     \\
5255835 & C & 7 &  77.15 $\pm$   1.31 &  7.766 &   4840 $\pm$  90 & -0.62 &  0.32 & 0.267 &  0.87 &  0.90 &  0.93 &  6.19 &  6.27 &  6.34 & 10.79 & 12.16 & 13.92     \\
5301647 & C & 7 & 137.05 $\pm$   2.33 & 12.294 &   4881 $\pm$ 100 & -0.70 &  0.22 & 0.311 &  0.85 &  0.89 &  0.91 &  4.57 &  4.64 &  4.67 & 10.56 & 11.42 & 13.59     \\
5306751 & C & 7 &  89.72 $\pm$   1.53 &  8.735 &   4838 $\pm$  87 & -0.52 &  0.21 & 0.164 &  0.93 &  0.94 &  0.95 &  5.87 &  5.90 &  5.94 &  9.92 & 10.77 & 11.36     \\
5445379 & C & 7 &  33.11 $\pm$   0.56 &  3.641 &   4754 $\pm$  80 & -0.65 &  0.28 & 0.488 &  1.43 &  1.46 &  1.52 & 12.15 & 12.26 & 12.43 &  1.92 &  2.20 &  2.40     \\
5602690 & C & 5 &  31.01 $\pm$   0.53 &  3.906 &   4615 $\pm$  78 & -0.58 &  0.23 & 0.257 &  0.88 &  0.91 &  0.96 &  9.88 &  9.96 & 10.15 &  9.84 & 12.11 & 12.98     \\
5696539 & C & 6 &  64.37 $\pm$   1.09 &  6.751 &   4811 $\pm$  79 & -0.57 &  0.25 & 0.434 &  0.85 &  0.86 &  0.94 &  6.73 &  6.76 &  6.99 & 10.25 & 14.60 & 15.20     \\
5812955 & C & 6 &  69.34 $\pm$   1.18 &  7.200 &   4803 $\pm$  78 & -0.59 &  0.21 & 0.449 &  1.03 &  1.04 &  1.06 &  6.94 &  6.97 &  7.01 &  6.48 &  6.94 &  7.37     \\
5858947 & C & 7 & 168.66 $\pm$   2.87 & 14.502 &   5002 $\pm$  87 & -0.83 &  0.26 & 0.749 &  0.87 &  0.89 &  0.91 &  4.13 &  4.16 &  4.20 &  9.97 & 11.03 & 11.83     \\
5953450 & A & 7 & 140.37 $\pm$   2.39 & 12.652 &   5054 $\pm$  96 & -0.67 &  0.04 & 0.996 &  0.90 &  0.94 &  0.96 &  4.58 &  4.64 &  4.68 &  8.25 &  8.97 & 10.21     \\
6363813 & C & 6 &  40.85 $\pm$   0.69 &  4.756 &   4748 $\pm$  81 & -0.66 &  0.29 & 0.273 &  0.89 &  0.91 &  0.93 &  8.64 &  8.71 &  8.78 & 10.28 & 11.23 & 12.31     \\
6547007 & C & 6 &  37.45 $\pm$   0.64 &  4.517 &   4728 $\pm$  73 & -0.70 &  0.20 & 0.275 &  0.80 &  0.80 &  0.90 &  8.57 &  8.60 &  8.95 & 11.09 & 16.63 & 17.35     \\
6752140 & C & 7 &  60.03 $\pm$   1.02 &  5.953 &   4764 $\pm$  74 & -0.53 &  0.31 & 0.214 &  1.18 &  1.20 &  1.22 &  8.19 &  8.23 &  8.29 &  4.29 &  4.65 &  4.98     \\
6837256 & C & 7 &  35.36 $\pm$   0.60 &  3.927 &   4731 $\pm$  73 & -0.57 &  0.24 & 0.279 &  1.22 &  1.24 &  1.28 & 10.83 & 10.89 & 11.00 &  3.55 &  3.97 &  4.19     \\
6924105 & C & 7 & 103.47 $\pm$   1.76 & 10.019 &   5123 $\pm$ 104 & -1.29 &  0.39 & 0.364 &  0.87 &  0.88 &  0.89 &  5.27 &  5.29 &  5.33 &  9.25 &  9.93 & 10.33     \\
7006903 & C & 7 &  44.75 $\pm$   0.76 &  5.148 &   4719 $\pm$  83 & -0.63 &  0.25 & 0.282 &  0.90 &  0.92 &  0.96 &  8.27 &  8.33 &  8.44 &  9.39 & 10.93 & 11.78     \\
7019157 & C & 5 &  27.47 $\pm$   0.47 &  3.438 &   4820 $\pm$  90 & -1.23 &  0.36 & 0.141 &  1.03 &  1.10 &  1.14 & 11.22 & 11.50 & 11.63 &  4.11 &  4.55 &  5.69     \\
7255651 & B & 6 &  22.48 $\pm$   0.38 &  3.004 &   4962 $\pm$  98 & -1.44 &  0.28 & 0.301 &  0.95 &  0.97 &  1.00 & 11.90 & 11.98 & 12.12 &  5.91 &  6.71 &  7.14     \\
7257193 & C & 4 &  15.89 $\pm$   0.27 &  2.322 &   4609 $\pm$  73 & -0.85 &  0.28 & 0.361 &  0.78 &  0.91 &  0.94 & 13.27 & 14.01 & 14.14 &  9.11 & 10.13 & 17.51     \\
7257814 & C & 5 &  30.26 $\pm$   0.51 &  3.741 &   4641 $\pm$  75 & -0.56 &  0.24 & 0.142 &  0.90 &  0.89 &  1.01 & 10.15 & 10.12 & 10.58 &  8.12 & 12.92 & 12.28     \\
7265189 & C & 6 &  85.02 $\pm$   1.45 &  8.449 &   4996 $\pm$  87 & -0.98 &  0.22 & 0.216 &  0.85 &  0.87 &  0.91 &  5.79 &  5.85 &  5.94 &  9.20 & 10.71 & 11.98     \\
7335713 & C & 6 & 158.37 $\pm$   2.69 & 13.732 &   4981 $\pm$  98 & -0.81 &  0.15 & 0.123 &  0.85 &  0.87 &  0.90 &  4.24 &  4.28 &  4.33 & 10.13 & 11.44 & 12.68     \\
7510918 & C & 6 &  31.99 $\pm$   0.54 &  4.011 &   4738 $\pm$  84 & -0.63 &  0.18 & 0.107 &  0.83 &  0.87 &  0.91 &  9.41 &  9.55 &  9.72 & 10.89 & 12.98 & 15.11     \\
7594337 & C & 6 &  97.12 $\pm$   1.65 &  9.233 &   4815 $\pm$  77 & -0.57 &  0.20 & 0.263 &  0.93 &  0.94 &  0.96 &  5.67 &  5.69 &  5.73 &  9.38 & 10.21 & 10.69     \\
7596219 & C & 4 &  19.35 $\pm$   0.33 &  2.699 &   4593 $\pm$  74 & -0.70 &  0.25 & 0.613 &  0.89 &  0.91 &  0.92 & 12.50 & 12.57 & 12.66 & 10.24 & 11.08 & 11.87     \\
7799904 & C & 7 &  77.60 $\pm$   1.32 &  7.820 &   4763 $\pm$  74 & -0.52 &  0.22 & 0.455 &  0.90 &  0.93 &  0.94 &  6.24 &  6.30 &  6.34 & 10.62 & 11.41 & 12.71     \\
7908109 & C & 5 &  52.38 $\pm$   0.89 &  5.802 &   4774 $\pm$  82 & -0.76 &  0.24 & 0.652 &  0.90 &  0.91 &  0.92 &  7.57 &  7.60 &  7.65 &  9.83 & 10.57 & 11.06     \\
8099102 & C & 5 &  26.61 $\pm$   0.45 &  3.390 &   4732 $\pm$  76 & -0.77 &  0.22 & 0.186 &  0.91 &  0.93 &  0.99 & 10.80 & 10.87 & 11.11 &  7.71 &  9.59 & 10.34     \\
8210058 & C & 6 &  39.97 $\pm$   0.68 &  4.762 &   4758 $\pm$  86 & -0.76 &  0.26 & 0.389 &  0.81 &  0.85 &  0.89 &  8.35 &  8.51 &  8.64 & 11.21 & 13.20 & 15.94     \\
8417027 & C & 6 &  31.41 $\pm$   0.53 &  3.880 &   4698 $\pm$  85 & -0.53 &  0.24 & 0.231 &  0.87 &  0.87 &  0.94 &  9.73 &  9.74 & 10.01 & 10.38 & 14.04 & 13.81     \\
8544630 & C & 6 &  73.96 $\pm$   1.26 &  7.495 &   4787 $\pm$  89 & -0.53 &  0.17 & 0.681 &  0.89 &  0.91 &  0.93 &  6.37 &  6.44 &  6.48 & 10.56 & 11.43 & 12.70     \\
8557457 & C & 6 &  33.98 $\pm$   0.58 &  4.094 &   4742 $\pm$  79 & -0.72 &  0.24 & 0.236 &  0.91 &  0.96 &  0.98 &  9.51 &  9.69 &  9.76 &  8.22 &  9.01 & 10.83     \\
8683589 & C & 4 &  16.58 $\pm$   0.28 &  2.328 &   4484 $\pm$  63 & -0.58 &  0.28 & 0.505 &  1.02 &  1.03 &  1.05 & 14.51 & 14.58 & 14.69 &  7.05 &  7.65 &  8.12     \\
8869235 & A & 6 &  58.51 $\pm$   0.99 &  6.318 &   5064 $\pm$ 116 & -1.03 &  0.23 & 0.984 &  0.88 &  0.89 &  0.91 &  7.06 &  7.10 &  7.15 &  9.19 &  9.92 & 10.45     \\
9072262 & C & 7 &  38.10 $\pm$   0.65 &  4.470 &   4708 $\pm$  72 & -0.66 &  0.21 & 0.149 &  0.86 &  0.90 &  0.92 &  8.80 &  8.96 &  9.04 & 10.48 & 11.44 & 13.75     \\
9084314 & B & 6 &  40.81 $\pm$   0.69 &  4.673 &   4787 $\pm$  82 & -0.56 &  0.08 & 0.079 &  0.91 &  0.94 &  0.97 &  8.73 &  8.83 &  8.92 &  8.66 &  9.66 & 10.93     \\
9206941 & C & 6 &  42.75 $\pm$   0.73 &  4.908 &   4761 $\pm$  79 & -0.53 &  0.19 & 0.207 &  0.93 &  0.95 &  0.97 &  8.56 &  8.61 &  8.67 &  9.32 & 10.11 & 10.87     \\
9292100 & C & 6 & 102.12 $\pm$   1.74 &  9.565 &   4885 $\pm$  97 & -0.54 &  0.26 & 0.912 &  0.94 &  0.96 &  0.98 &  5.54 &  5.58 &  5.62 &  8.99 & 10.01 & 10.74     \\
9339711 & B & 5 &  20.23 $\pm$   0.34 &  2.815 &   4908 $\pm$  84 & -1.68 &  0.29 & 0.592 &  0.84 &  0.87 &  0.91 & 11.97 & 12.13 & 12.32 &  8.00 &  9.36 & 10.61     \\
9544002 & C & 6 &  44.05 $\pm$   0.75 &  4.942 &   4705 $\pm$  79 & -0.64 &  0.26 & 0.094 &  0.91 &  0.94 &  0.97 &  8.39 &  8.48 &  8.58 &  8.87 & 10.17 & 11.21     \\
9637337 & C & 6 &  63.82 $\pm$   1.08 &  6.804 &   4917 $\pm$  83 & -1.26 &  0.23 & 0.540 &  0.83 &  0.84 &  0.86 &  6.62 &  6.66 &  6.70 & 10.55 & 11.22 & 11.89     \\
9696716 & B & 4 &  24.30 $\pm$   0.41 &  3.294 &   4952 $\pm$  94 & -1.60 &  0.23 & 0.662 &  0.88 &  0.90 &  0.93 & 11.02 & 11.11 & 11.26 &  7.34 &  8.37 &  9.06     \\
9773214 & C & 6 &  43.22 $\pm$   0.73 &  4.947 &   4729 $\pm$  73 & -0.51 &  0.11 & 0.305 &  0.94 &  0.95 &  0.96 &  8.55 &  8.58 &  8.62 &  9.48 & 10.05 & 10.57     \\
9777293 & C & 7 & 181.27 $\pm$   3.08 & 15.038 &   4863 $\pm$  79 & -0.50 &  0.23 & 0.248 &  0.95 &  0.96 &  0.98 &  4.15 &  4.17 &  4.19 &  9.21 &  9.92 & 10.59     \\
9777355 & C & 7 & 197.25 $\pm$   3.35 & 15.992 &   4914 $\pm$  81 & -0.50 &  0.26 & 0.254 &  0.91 &  0.93 &  0.94 &  3.93 &  3.96 &  3.98 & 10.50 & 11.52 & 12.33     \\
9908656 & C & 4 &  15.80 $\pm$   0.27 &  2.371 &   4543 $\pm$  68 & -0.59 &  0.19 & 0.293 &  0.78 &  0.81 &  0.85 & 13.27 & 13.47 & 13.69 & 14.26 & 17.05 & 19.69     \\
9968040 & C & 7 &  43.58 $\pm$   0.74 &  5.002 &   4690 $\pm$  71 & -0.60 &  0.25 & 0.397 &  0.90 &  0.91 &  0.92 &  8.41 &  8.44 &  8.49 & 11.05 & 11.80 & 12.46     \\
10001167 & C & 4 &  19.68 $\pm$   0.33 &  2.734 &   4556 $\pm$  66 & -0.61 &  0.25 & 0.436 &  0.92 &  0.94 &  0.96 & 12.64 & 12.73 & 12.83 &  9.38 & 10.33 & 11.14     \\
10001440 & C & 6 &  44.12 $\pm$   0.75 &  5.106 &   4712 $\pm$  79 & -0.61 &  0.26 & 0.269 &  0.91 &  0.92 &  0.93 &  8.31 &  8.35 &  8.39 & 10.31 & 11.01 & 11.68     \\
10083815 & A & 5 &  17.86 $\pm$   0.30 &  2.559 &   4700 $\pm$  78 & -0.94 &  0.19 & 0.953 &  0.89 &  0.90 &  0.93 & 12.84 & 12.94 & 13.08 &  8.50 &  9.64 & 10.37     \\
10196630 & C & 6 &  23.89 $\pm$   0.41 &  3.174 &   4671 $\pm$  71 & -0.77 &  0.28 & 0.297 &  0.84 &  0.85 &  0.87 & 11.04 & 11.11 & 11.20 & 12.12 & 13.46 & 14.34     \\
10200377 & C & 7 & 141.42 $\pm$   2.40 & 12.464 &   4837 $\pm$  78 & -0.52 &  0.21 & 0.089 &  0.92 &  0.94 &  0.95 &  4.65 &  4.68 &  4.70 &  9.90 & 10.66 & 11.39     \\
10318430 & C & 7 & 154.19 $\pm$   2.62 & 13.000 &   4987 $\pm$  86 & -0.72 &  0.25 & 0.487 &  1.00 &  1.02 &  1.04 &  4.66 &  4.68 &  4.71 &  6.43 &  6.94 &  7.40     \\
10318787 & C & 7 &  42.40 $\pm$   0.72 &  5.097 &   5014 $\pm$ 105 & -1.46 &  0.29 & 0.550 &  0.86 &  0.88 &  0.89 &  8.14 &  8.20 &  8.24 &  8.73 &  9.18 &  9.84     \\
10423340 & C & 6 & 226.86 $\pm$   3.86 & 18.865 &   5010 $\pm$  87 & -0.87 &  0.22 & 0.350 &  0.82 &  0.85 &  0.87 &  3.41 &  3.45 &  3.48 & 11.34 & 12.38 & 13.83     \\
10449862 & C & 6 &  78.18 $\pm$   1.33 &  7.808 &   4816 $\pm$  77 & -0.61 &  0.26 & 0.963 &  0.92 &  0.93 &  0.95 &  6.28 &  6.31 &  6.35 &  9.81 & 10.66 & 11.39     \\
10451228 & C & 6 &  42.90 $\pm$   0.73 &  4.961 &   4768 $\pm$  75 & -0.65 &  0.24 & 0.182 &  0.92 &  0.93 &  0.94 &  8.46 &  8.50 &  8.54 &  9.67 & 10.31 & 10.94     \\
10796857 & C & 7 &  88.02 $\pm$   1.50 &  7.773 &   5072 $\pm$  90 & -0.71 &  0.26 & 0.827 &  1.48 &  1.51 &  1.54 &  7.48 &  7.53 &  7.59 &  1.71 &  1.86 &  1.98     \\
10857579 & C & 5 &  31.41 $\pm$   0.53 &  3.812 &   4659 $\pm$  85 & -0.66 &  0.21 & 0.592 &  0.95 &  1.00 &  1.02 & 10.15 & 10.33 & 10.39 &  7.32 &  7.82 &  9.57     \\
11014721 & C & 7 &  56.03 $\pm$   0.95 &  6.066 &   4782 $\pm$  84 & -0.57 &  0.12 & 0.333 &  0.92 &  0.93 &  0.95 &  7.39 &  7.42 &  7.48 &  9.39 & 10.38 & 10.91     \\
11037292 & C & 5 &  16.90 $\pm$   0.29 &  2.405 &   4485 $\pm$  63 & -0.62 &  0.27 & 0.617 &  0.93 &  0.95 &  0.96 & 13.67 & 13.78 & 13.87 &  9.44 & 10.16 & 11.07     \\
11044668 & C & 7 &  50.05 $\pm$   0.85 &  5.671 &   4815 $\pm$  79 & -0.86 &  0.27 & 0.378 &  0.83 &  0.84 &  0.87 &  7.54 &  7.58 &  7.65 & 11.71 & 12.99 & 13.75     \\
11046830 & C & 7 &  56.69 $\pm$   0.96 &  6.161 &   4745 $\pm$  86 & -0.78 &  0.19 & 0.441 &  0.91 &  0.93 &  0.96 &  7.38 &  7.42 &  7.51 &  8.26 &  9.38 & 10.02     \\
11181828 & B & 7 &  33.44 $\pm$   0.57 &  4.098 &   4790 $\pm$  83 & -0.92 &  0.21 & 0.247 &  0.93 &  0.97 &  1.00 &  9.69 &  9.84 &  9.95 &  6.69 &  7.60 &  8.81     \\
11200680 & C & 6 &  38.86 $\pm$   0.66 &  4.545 &   4763 $\pm$  76 & -0.72 &  0.30 & 0.275 &  0.92 &  0.94 &  0.96 &  8.97 &  9.04 &  9.10 &  8.96 &  9.75 & 10.62     \\
11230715 & C & 7 & 185.18 $\pm$   3.15 & 15.303 &   4958 $\pm$ 100 & -0.56 &  0.24 & 0.475 &  0.90 &  0.93 &  0.97 &  4.01 &  4.06 &  4.12 &  9.15 & 10.76 & 12.22     \\
11289675 & C & 5 &  26.00 $\pm$   0.44 &  3.422 &   4676 $\pm$  80 & -0.68 &  0.26 & 0.045 &  0.90 &  0.91 &  0.94 & 10.81 & 10.84 & 10.97 &  9.46 & 11.02 & 11.09     \\
11296084 & C & 4 &  16.59 $\pm$   0.28 &  2.351 &   4465 $\pm$  62 & -0.55 &  0.23 & 0.059 &  0.92 &  0.95 &  0.97 & 13.94 & 14.11 & 14.23 &  9.39 & 10.39 & 11.67     \\
11445495 & C & 5 & 204.83 $\pm$   3.48 & 16.938 &   4973 $\pm$  85 & -0.81 &  0.29 & 0.665 &  0.82 &  0.84 &  0.91 &  3.66 &  3.70 &  3.79 & 10.31 & 13.42 & 14.79     \\
11498043 & C & 6 &  32.87 $\pm$   0.56 &  4.042 &   4740 $\pm$  80 & -0.79 &  0.29 & 0.027 &  0.82 &  0.85 &  0.91 &  9.22 &  9.32 &  9.56 & 10.48 & 13.37 & 15.07     \\
11509608 & C & 5 &  33.40 $\pm$   0.57 &  4.057 &   4733 $\pm$  73 & -0.78 &  0.27 & 0.501 &  0.89 &  0.93 &  0.95 &  9.61 &  9.74 &  9.83 &  8.82 &  9.74 & 11.19     \\
11598459 & C & 7 & 124.02 $\pm$   2.11 & 11.266 &   4884 $\pm$  87 & -0.58 &  0.31 & 0.534 &  0.85 &  0.87 &  0.90 &  4.83 &  4.86 &  4.92 & 12.25 & 14.19 & 15.21     \\
11717920 & C & 6 &  22.41 $\pm$   0.38 &  2.807 &   4670 $\pm$  84 & -0.62 &  0.22 & 0.311 &  1.29 &  1.31 &  1.35 & 13.94 & 14.04 & 14.18 &  2.82 &  3.17 &  3.29     \\
11752358 & C & 6 &  64.59 $\pm$   1.10 &  6.503 &   4946 $\pm$  92 & -0.74 &  0.17 & 0.138 &  1.02 &  1.04 &  1.06 &  7.32 &  7.37 &  7.42 &  5.73 &  6.19 &  6.64     \\
11753104 & C & 6 &  33.40 $\pm$   0.57 &  3.866 &   4727 $\pm$  79 & -0.60 &  0.26 & 0.281 &  1.11 &  1.14 &  1.18 & 10.66 & 10.77 & 10.90 &  4.57 &  5.22 &  5.81     \\
11968543 & C & 6 & 116.35 $\pm$   1.98 & 10.869 &   4976 $\pm$  90 & -0.95 &  0.24 & 0.536 &  0.87 &  0.88 &  0.90 &  4.99 &  5.01 &  5.05 &  9.81 & 10.56 & 11.14     \\
12111110 & A & 6 &  29.48 $\pm$   0.50 &  3.773 &   4890 $\pm$  99 & -1.10 &  0.13 & 0.990 &  0.88 &  0.87 &  1.07 & 10.13 & 10.09 & 10.84 &  4.95 & 10.09 &  9.73     \\
12115227 & C & 6 &  97.33 $\pm$   1.65 &  9.387 &   4827 $\pm$  77 & -0.62 &  0.26 & 0.311 &  0.92 &  0.94 &  0.96 &  5.60 &  5.64 &  5.68 &  9.28 & 10.22 & 11.03     \\
12207740 & B & 6 &  38.67 $\pm$   0.66 &  4.601 &   4996 $\pm$  95 & -1.22 &  0.32 & 0.251 &  0.88 &  0.92 &  0.94 &  8.77 &  8.91 &  8.99 &  7.78 &  8.56 & 10.00     \\
12207840 & C & 5 &  77.12 $\pm$   1.31 &  7.837 &   4802 $\pm$  76 & -0.53 &  0.19 & 0.264 &  0.92 &  0.94 &  0.96 &  6.31 &  6.36 &  6.39 &  9.70 & 10.34 & 11.32     \\
12253381 & A & 5 &  21.87 $\pm$   0.37 &  3.028 &   5002 $\pm$ 100 & -1.71 &  0.07 & 0.963 &  0.87 &  0.89 &  0.92 & 11.61 & 11.70 & 11.84 &  7.55 &  8.51 &  9.20     \\
12300740 & C & 6 &  31.70 $\pm$   0.54 &  3.657 &   4778 $\pm$  76 & -0.64 &  0.19 & 0.309 &  1.28 &  1.30 &  1.34 & 11.69 & 11.75 & 11.86 &  2.87 &  3.16 &  3.34     \\

\hline

\end{longtable}
\endgroup
\end{landscape}